\documentclass{article}

\usepackage{arxiv}

\usepackage[utf8]{inputenc} 
\usepackage[T1]{fontenc}    
\usepackage{hyperref}       
\usepackage{url}            
\usepackage{amsfonts}       
\usepackage{xcolor}
\usepackage{graphicx}

\title{Stellar physics with high-resolution UV spectropolarimetry}

\author{
  Contact scientist: {\Large Julien Morin}\\
  Laboratoire Univers et Particules de Montpellier (LUPM)\\
  Universit\'e de Montpellier, CNRS\\
  34095 Montpellier, France\\
  \texttt{julien.morin@umontpellier.fr} \\
  }

\begin{document}
\maketitle

\pagebreak
\begin{abstract}
Current burning issues in stellar physics, for both hot and cool stars, concern their magnetism. In hot stars, stable magnetic fields of fossil origin impact their stellar structure and circumstellar environment, with a likely major role in stellar evolution. However, this role is complex and thus poorly understood as of today. It needs to be quantified with high-resolution UV spectropolarimetric measurements. In cool stars, UV spectropolarimetry would provide access to the structure and magnetic field of the very dynamic upper stellar atmosphere, providing key data for new progress to be made on the role of magnetic fields in heating the upper atmospheres, launching stellar winds, and more generally in the interaction of cool stars with their environment (circumstellar disk, planets) along their whole evolution. UV spectropolarimetry is proposed on missions of various sizes and scopes, from POLLUX on the 15-m telescope LUVOIR to the Arago M-size mission dedicated to UV spectropolarimetry. 
\end{abstract}

\section{Scientific Context}
Stars form from material in the interstellar medium (ISM). As they accrete matter from their parent molecular cloud, planets can also form. During the formation and throughout the entire life of stars and planets, a few key basic physical processes, involving in particular magnetic fields, winds, rotation, and binarity, directly affect the internal structure of stars, their dynamics, and immediate circumstellar environment. They consequently drive stellar evolution, but also fundamentally impact the formation, environments, and fate of planets. 
Here we argue that enabling high-resolution spectropolarimetry at UV wavelengths is mandatory to get access to very powerful diagnostics to study the formation, evolution, and 3D dynamical environments of stars, and their role on the formation and evolution of planets and life. 

The UV domain is crucial in stellar physics, as it is particularly rich in atomic and molecular transitions, and
covers the region in which the intrinsic spectral energy distributions of hot stars peak. It contains forest of lines of different species, including some that are exclusively found in the UV part of stellar spectra, and it is thus most useful, e.g., for quantitative determinations of chemical abundances. The lower energy levels of these lines are less likely to depopulate in low density environments such as chromospheres, circumstellar shells, stellar winds, nebulae, and the ISM, and so remain the only useful plasma diagnostics in most of these environments.\\
Moreover, the UV spectrum is extremely sensitive to the presence of small amounts of hot gas in dominantly cool environments. This allows the detection and monitoring of various phenomena: accretion continua in young stars, magnetic activity, chromospheric heating, coronae, plages and faculae on cool stars, and intrinsically faint, but hot, companions of cool stars. The UV domain is also that in which Sun-like stars exhibit their greatest potential hostility (or not) to Earth-like life, population III stars must have shone the brightest, conversion of kinetic energy into radiation by accretion processes most strongly impacts stellar formation and evolution, and the “Fe curtain” features respond to changes in local irradiation. Moreover, many light scattering and polarising processes are stronger at UV wavelengths.\\
In addition, most cool stars and a fraction of hot stars are magnetic, and their magnetic field interacts with
their wind and environment, modifies their structure and surface abundances, and contributes to the transport of
angular momentum. 

With spectropolarimetry, one can address, with unprecedented detail, these important issues in
stellar physics, from stellar magnetic fields to surface inhomogeneities, surface differential rotation to activity cycles and magnetic braking, from microscopic diffusion to turbulence, convection and circulation in stellar interiors, from abundances in stellar atmospheres to stellar winds and accretion discs, from the early phases of stellar formation to the late stages of evolution, from extended circumstellar environments to the distant ISM. However, measuring polarisation directly in UV wind-sensitive lines has never been done, and will be extremely useful to trace polarisation along field lines. Finally, the scope of polarimetry also includes linear polarisation and depolarisation processes in circumstellar environments, e.g., in accretion or decretion discs, or from exoplanets.

\section{Massive and hot stars}

Massive stars are those with an initial mass above 8 solar masses. They spend most of their lives as O or B stars on the main sequence, then they evolve after a few million years either to the Red Supergiant (RSG) phase at lower masses, or in some if not all cases to Luminous Blue Variables (LBVs) and then the Wolf-Rayet (WR) state for the most massive objects. Massive stars  provide heavy chemical elements to the Universe and dominate the interstellar radiation field. Moreover, they are the progenitors of  core-collapse supernovae and gamma-ray bursts, leaving behind compact objects such as neutron stars and black holes. These, when in binary systems, may trigger the emission of gravitational waves during coalescence. In addition, due to their luminosity and spectroscopic features, the successive phases of massive stars and starbursts can be observed out to large distances. Therefore, they are essential for many domains of astrophysics, such as stellar and planetary formation and galactic structure and evolution. It is therefore crucial to understand the physical processes at work in massive stars.

It is common knowledge that mass is the prime parameter governing the structure and evolution of a star. A second parameter is its chemical composition, often parametrized through its heavy-element content (also known as metallicity). In the case of massive stars, the evolution is also strongly impacted by mass loss through stellar winds \cite{vink01}, rotation, and binary interactions \cite{langer12}. These processes are strongly coupled, making the prediction of massive star evolution a very difficult task \cite{groh13}.\\
Very recently, thanks to high precision photometric missions such as TESS, asteroseismology has also provided new insights into the interior of those stars. However, one physical ingredient remains very poorly understood in massive stars and hot stars in general: their magnetic field, and its impact on their structure, evolution, and environment.
 As a consequence, no global consensus has been achieved yet, on the details of the evolution and properties of massive stars.

About 10\% of O, B, and A stars host a magnetic field of fossil origin, usually dipolar but inclined with respect to the stellar rotation axis, with a polar field strength ranging from a few hundreds to a few ten thousands Gauss (\cite{neiner2015}; \cite{grunhut15}). The $\sim$90\% of stars that do not host such a field may nevertheless host an ultra-weak field of the order of 1 Gauss, such as those recently discovered in some A and Am stars \cite{blazere2016}. 

Since the fossil fields of hot main sequence stars are long-lived, and organized on global scales, they are expected to be highly influential in the context of stellar structure and evolution. In recent years, modern models of stellar evolution have predicted that magnetic fields have a critical impact on stellar evolution (\cite{maeder03} and subsequent papers in the series, \cite{song16}). Magnetic fields are predicted to be responsible for modification of convective and circulatory interior flows, redistribution of angular momentum and nucleosynthetic chemicals, channeling and modification of mass loss, and shedding of rotational angular momentum through magnetic braking (\cite{mestel99}). Ultimately, these effects lead to important modification of stellar evolutionary pathways and stellar feedback effects, such as mechanical energy deposition in the ISM and supernova explosions \cite{heger05}, and hence the properties of stellar remnants and potentially the structure and chemistry of the local Galactic environment. It is therefore remarkable that modern stellar evolution models are only just beginning to consider including magnetic fields in a realistic manner.

The basic consequences of magnetic fields for stellar evolution fall into two general categories: (i) interaction of interior fields with interior fluid motions, impacting the internal rotational profile, angular momentum, and chemical transport (e.g. \cite{mestel99,mathis05}); and (ii) interaction of surface fields with the stellar wind, leading to magnetic braking of surface layers and reduction of the surface mass-loss rate (\cite{uddoula02}, \cite{uddoula08}, \cite{uddoula09}, \cite{meynet11}). For example, studies of magnetic Herbig Ae/Be and Ap/Bp stars have shown that their magnetic field brakes their rotation rate during the early phases of their life, and evidence for spindown on the MS has also been found in hotter stars \cite{townsend10}. Recent studies also suggest, and in some cases demand, that magnetic fields have direct and ubiquitous consequences for evolution. For example, in order to explain the post-MS gap of blue supergiants, \cite{petermann15} have proposed that these objects evolve from magnetic MS stars. In addition, \cite{maeder14} have examined the role of strong, organized fields in the cores of red supergiants, with implications for the general spin rates of (magnetic) white dwarfs and pulsars.

\begin{figure*}[t]
    \centering
    \resizebox{\hsize}{!}{\includegraphics{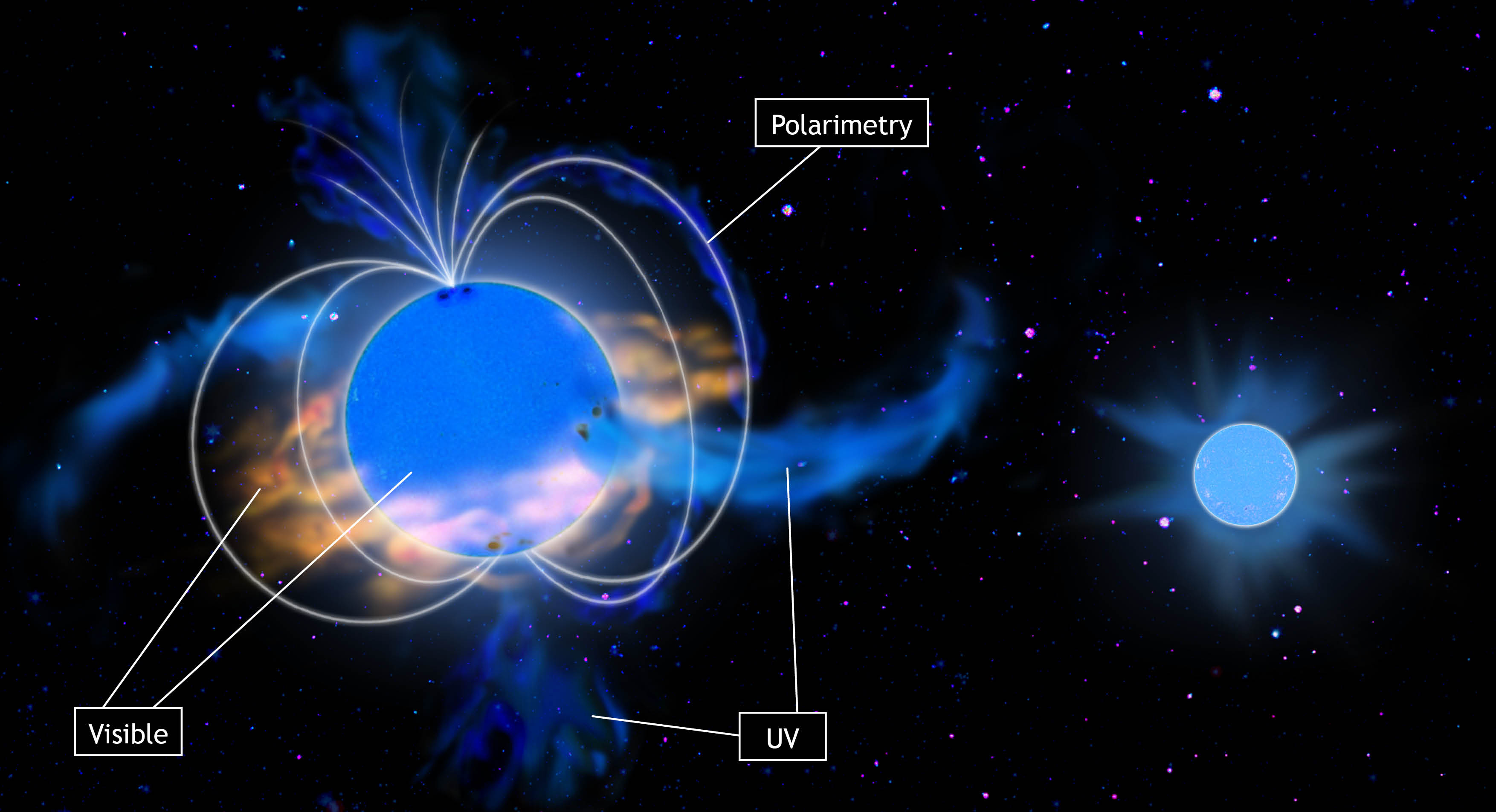}}
    \caption{Schematic view of a hot star with its fossil magnetic field lines, channeled polar wind, surface spots, equatorial magnetosphere, corotating interaction regions, and a stellar companion. Copyright: S. Cnudde.}
    \label{fig:hotsketch}
\end{figure*}

The current burning issues about massive stars therefore concern their magnetism and are the following:
\begin{itemize}
    \item {\it Why are there only $\sim$10\% of magnetic hot stars with a field above a few hundreds of Gauss?} Since the magnetic fields are of fossil origin, i.e. descendants from a seed field present in the molecular cloud from which the star was formed and enhanced by a dynamo during the early phases of the life of the star (when it was fully convective), the occurrence rate of magnetic fields is likely related to the initial conditions of stellar formation. Quantitatively relating the initial conditions to the presence of a stable field would give us important insight into the early phases of stellar evolution.
    \item {\it Why are there less magnetic hot stars in short-period binary systems ($\sim$2\%; \cite{alecian15}) than among single stars ($\sim$10\%)?} This is also probably related to the fossil origin of magnetic fields in hot stars, possibly to the difficulty to fragment cores in a magnetized medium \cite{commercon11}. Understanding this occurrence rate difference between single and binary hot stars would cast new light on star formation and binary formation.
    \item {\it How do fossil magnetic fields evolve and how do they impact the evolution of the star during and after the main sequence? Is magnetic flux conserved throughout the life of the star or are there some processes at work producing a decay or an enhancement of the field during stellar evolution?}
    The commonly assumed model of magnetic flux conservation (with magnetic field strength decreasing at the stellar surface only because of the increase of the stellar radius) has been recently debated (e.g. \cite{blazere15}, \cite{fossati16}, especially for the most massive stars (\cite{shultz19}), and as such theoretical predictions are prone to embarrassing uncertainties when it comes to such important objects as massive stars.
    \item {\it Do dynamo fields develop in the convective regions that appear in the radiative envelope in the second part of the stellar life? If so, what is the result of the interaction between the fossil field and the dynamo field(s) on stellar structure and evolution?} 
    As magnetic fields influence stellar evolution, so are magnetic fields expected to transform in response to changes in the structure of the stars in which they are embedded. In particular, as hot stars age, their radius expands dramatically and convective regions develop in their radiative envelope. Local dynamos probably develop in these convective zones, providing an opportunity to study the unique interactions between the post-main sequence dynamo and the pre-existing fossil field. In particular, it is expected that the dynamo-fossil interaction could enhance the local dynamo fields and modify the configuration of the global fossil field (\cite{featherstone09}, \cite{auriere08}) but statistics on evolved magnetic hot stars are too poor to confirm this prediction observationally so far (\cite{neiner18}). 
    \item {\it How do magnetic fields interact with fluid motions inside the star, impact the internal rotation profile, the transport of chemical elements and of angular momentum, and thus impact stellar evolution?} The inclusion of magnetic field (braking) in stellar evolution yields very different results for the surface abundances, depending on the assumed rotation law inside the star (\cite{meynet11}). For stars with differential rotation, the net result of magnetic braking is to make mixing faster and stronger. However, when a star hosts a fossil field of a few Gauss or more, solid-body rotation settles in, internal mixing is inhibited and, of the field is strong enough, the star spins down rapidly. In this case, surface abundance ratios are lower than in models without magnetic braking. Overcoming the basic limitation of this initial study, namely that the surface magnetic field strength remains constant during stellar evolution, \cite{keszthelyi19} showed that stars having evolving dipolar surface fossil magnetic field are expected to have low surface rotation and high surface nitrogen enhancement already on the main sequence. On the other hand, their results show that magnetic braking enhances the chemical enrichment if the star undergoes radial differential rotation, consistently with the results of \cite{meynet11}. Conversely, the enrichment is reduced if the star rotates as a solid body. This study, however, does not take the impact of the magnetic field on the internal structure of the star into account. Clearly, to which extent magnetic fields play a role in surface chemical enrichment is an open question. It further relates to prescriptions for rotation-related quantities such as the mixing-length, angular momentum transport processes, and overshooting, which impact the global structure of a massive star, altering its evolution and final fate (\cite{martins13}), thus questioning the robustness of predictions of the evolution of massive stars. 
    
    \begin{figure*}[t]
    \centering
    \resizebox{\hsize}{!}{\includegraphics{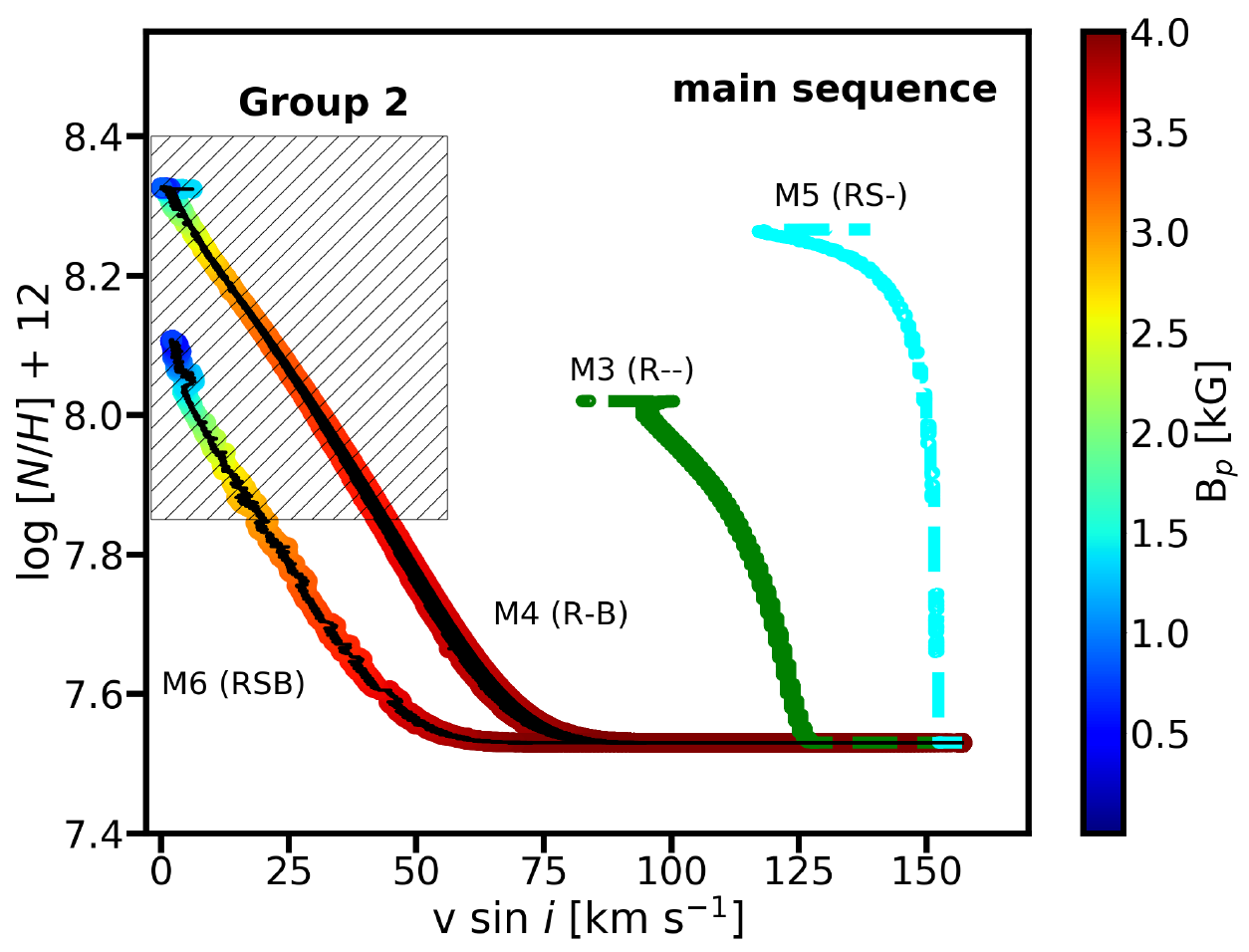}}
    \caption{Hunter diagram (nitrogen abundance versus $\it v$ sin $\it i$)
of the main sequence phase of rotating models where evolution of the surface magnetic
field is accounted for. Group 2 stars are slow rotators that have notable surface enrichment. They are denoted with the hatched box on the diagram and the polar magnetic field strength is colour-coded. The evolution of the models begins at the lower right corner of the diagram.
From Keszthelyi et al. (2019).
}
    \label{fig:keszthelyi}
\end{figure*}

    \item {\it How do magnetic fields impact the mass-loss rate of hot stars, in particular through the confinement of the wind particles into magnetospheres around the stars and magnetic braking (e.g. \cite{meynet11})?} It has been known for some time now that a surface dipolar magnetic field of sufficient strength (typically greater than about 1~kG) can considerably affect the winds from a hot massive star by confining part of the wind and preventing it from escaping the star (e.g. \cite{uddoula08}; \cite{bard16}). More precisely, the wind ionised outflow is channeled along the (dipolar) magnetic field lines down to the magnetic equator where it forms a disk. Part of the outflowing matter then falls back onto the star, lowering the net mass loss. This magnetic confinement explains why magnetic fields actually control the mass loss of massive stars, although the reduction is known only approximately from 2D and (limited) 3D MHD models of magnetized winds (e.g. \cite{uddoula02}, \cite{uddoula13}). Implementing the quenching of the mass loss produced by a surface dipolar magnetic field, \cite{petit17} and \cite{georgy17} showed that this phenomenon allows the star to maintain a higher mass during its evolution, respectively in the context of "heavy" stellar-mass black holes (with masses $>$ 25 M$_\odot$) such as those whose merger was reported by LIGO (\cite{abbott16, abbott17a, abbott17b}), and in the context of pair-instability supernovae.
    
        \begin{figure*}[t]
    \centering
    \resizebox{\hsize}{!}{\includegraphics{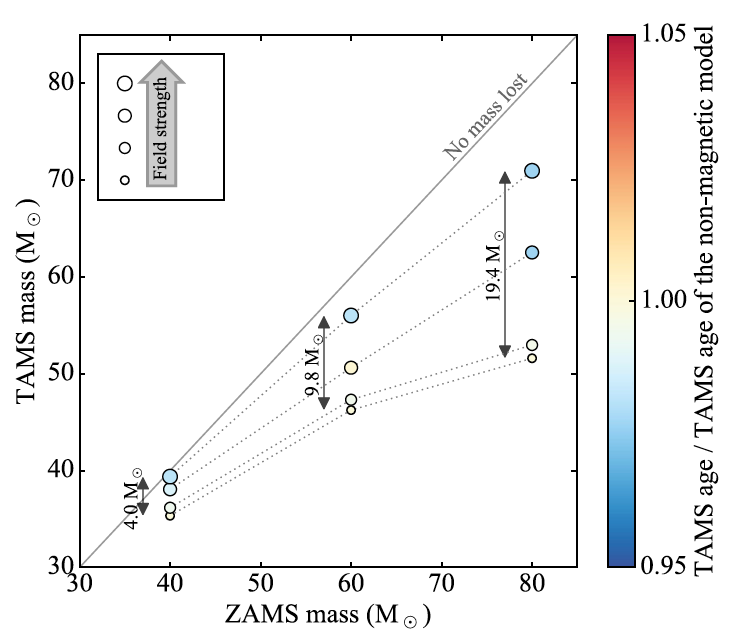}}
    \caption{Mass at the TAMS as a function of the initial mass at the ZAMS.
The initial field strength increases with increased symbol size. The points
are coloured according to the MS lifetime of the model as compared to the
non-magnetic model of the same initial mass. The vertical arrows indicate
the numerical value of the difference in TAMS stellar mass between the
most magnetic and non-magnetic models.
From Petit et al. (2017).
}
    \label{fig:petit}
\end{figure*}

    Another consequence is that during main sequence and post-main sequence evolution, fossil magnetic fields couple strongly to stellar winds, enhancing the shedding of rotational angular momentum through magnetic braking (\cite{townsend10}). 
    Investigating how magnetic fields couple to stellar winds and impact rotation, the intensity of the mass-loss, as well as the global structure of the wind is a critical issue, since the present uncertainties restrict our understanding of the initial-to-final mass relation, the prediction of final fates, and the suitability of these stars as progenitors of heavy mass black holes, long-soft gamma ray-bursts, and pair-instability supernovae. 
    \item {\it How do main sequence  fossil fields transform into the fields observed in stellar remnants?} An important objective of stellar magnetism research is to understand the evolution of magnetic fields across the entire span of stellar lives, from star formation to stellar remnants. White dwarfs and neutron stars represent the final stellar objects in which fields can be probed. The characteristics of magnetic fields in these objects (e.g. \cite{ferrario15}) have historically been interpreted as having a fossil origin, often directly connected through stellar evolution to their main sequence progenitors (e.g. \cite{wickramasinghe05}). However, how main sequence fossil fields transform into the fields observed in stellar remnants is unknown. More recently, binary merger/common envelope models of white dwarf magnetism have been proposed and examined (\cite{tout08, briggs15}), and concurrent dynamo contributions have been proposed (\cite{valyavin14} and references therein). An important key to making progress understanding stellar remnants is better characterization of the magnetic fields of giant and supergiant stellar populations, to understand the missing link between the known MS magnetic fields and fields in remnants.
    \item {\it How do magnetic fields contribute to the energy budget and type of supernovae explosions (\cite{guilet10}) and modify the very end stages of stellar evolution (e.g. formation of magnetars)?} Magnetic fields impact the death of hot stars directly and indirectly. First, the magnetic field in the deep interior (inner 2 M$_{\odot}$) of the star can directly contribute to the energy budget of the supernova explosion, leading to a bipolar explosion (\cite{burrows07}). Magneto-rotational interactions can amplify the magnetic field during the final stages of evolution, leading to the birth of magnetars as possible sources of long gamma-ray bursts (LGRBs) and superluminous supernovae (\cite{mosta15}). Secondly, magnetic fields impact the stellar death indirectly through their effect on convection and on the internal rotation profile. In particular, the turbulent kinetic energy associated to convection inhomogeneities in the oxygen and silicon layers may have a decisive influence on the supernova explosion scenario (\cite{couch13, couch15}, \cite{mueller15}). Moreover, the rotation profile in the stably stratified iron core and the convective oxygen and silicon layers is instrumental for the birth properties of pulsars: their angular momentum is redistributed by the one-armed spiral of the standing accretion shock instability (SASI) during the first second of the explosion (\cite{blondin07}, \cite{yamasaki08}). The rotational energy can thus directly impact the explosion scenario through the generation of a one-armed spiral dominated by the SASI (\cite{iwakami14}, \cite{nakamura14}) and the corotation instability (aka low T/W instability in \cite{takiwaki16}). As a consequence, including the impact of magnetic fields on the energy budget of the explosion, stellar convection, and rotation profile in supernova explosion models is necessary to understand the final stages of hot stars.
\end{itemize}

Since massive stars emit most of their radiation in the UV domain, and show atomic and molecular lines coming from the photosphere and wind in this wavelength range, these important topics would be best studied with a high-resolution UV spectropolarimeter. UV spectroscopy would indeed allow to characterize the wind and mass-loss properties of hot stars, while the UV polarimetric capabilities would provide, for the first time, a 3D mapping of the magnetized environment. Such maps could be performed on stars with various parameters (rotation, mass,...) and age to understand the impact of magnetic fields along stellar evolution. For those UV spectropolarimetric measurements, high spectral resolution is needed (R$\geq$30000) to allow for sufficient spatial resolution in the 3D maps.\\
Moreover, if the UV spectropolarimeter is placed on a sufficiently large telescope, hot stars outside our Galaxy, such as those in the Magellanic Clouds, could be measured as well providing the first insights into magnetic properties of stars in the metal-poor regime, opening a new window to quantitative stellar physics at metallicities corresponding to the peak of star formation in the Universe. Moreover, in light of new results from \cite{petit17} where magnetic fields can strongly quench mass-loss, detecting and quantifying magnetic fields in metal-poor environments becomes even more interesting, in the context of how they impact on the evolutionary end products (neutron stars, magnetars and the associated LGRBs, black holes). 

\section{Cool stars and their environments}
\subsection{UV spectroscopy and spectropolarimetry of cool stars}

Although cool stars emit only a small fraction of their bolometric luminosity at UV wavelengths and shorter, the UV spectra of these objects are extremely rich in both atomic and molecular lines. These lines are essential to determine the chemical composition of stars \cite{asplund04}, a fundamental ingredient in our understanding of the formation, structure, and evolution of stars and of their planetary systems, and of the chemical evolution of the Galaxy. A specificity of the UV emission of cool stars lies in the fact that a large fraction of the observed UV flux originates from the magnetically-heated upper atmosphere rather than the photosphere. The UV spectra of cool stars therefore represent a unique source of information on their magnetism, upper atmosphere and the way they interact with their environment, complementary with observations at optical wavelengths, in particular.

As opposed to the case of hotter stars discussed in the previous section, the magnetic fields of cool stars are thought to be generated through dynamo action: the complex interplay between convection and rotation results in the generation of time-dependent magnetic fields structured on a range of spatial scales with properties depending on stellar parameters such as mass and rotation. These magnetic fields and the resulting activity phenomena play a key role in the physics of cool stars and their planetary systems along their whole evolution \cite{mestel05}: from the the magnetospheric accretion of matter on the protostar and the conditions in the protoplanetary disk where planets form,  to the spin evolution along the whole stellar evolution, the high-energy radiation setting the habitability of orbiting planets, and until the final stages of evolution with a proposed role in mass-loss processes and thus in the chemical enrichment of the ISM.

With high-resolution spectroscopy and spectropolarimetry at visible and near-IR wavelengths we are now starting to get an overview of the main properties of the photospheric magnetic fields of cool stars along their evolution from young T~Tauri stars to evolved giants \cite{donati09}. The next step with combined UV/visible spectroscopy and spectropolarimetry consists in establishing quantitatively how these magnetic fields impact the environment of the star -- from its upper atmosphere, to the circumstellar disk (for young stars) and orbiting planets.

\begin{figure*}[t]
    \centering
    \resizebox{\hsize}{!}{\includegraphics{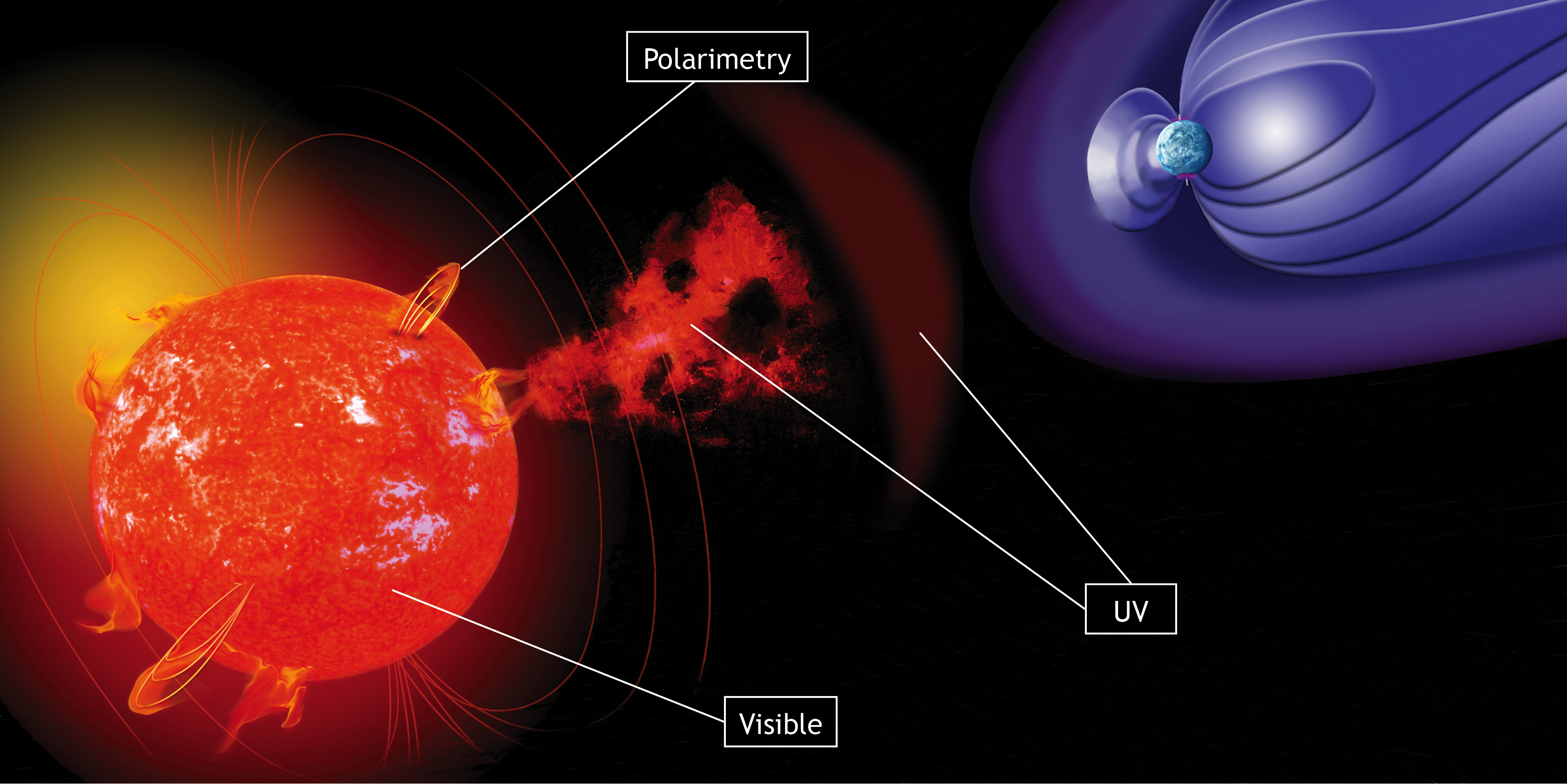}}
    \caption{Schematic view of a cool star with its dynamo magnetic field, surface faculae and plages, wind, a coronal mass ejection, and a bow shock between the star and its planet. Copyright: S.~Cnudde.}
    \label{fig:coolsketch}
\end{figure*}

\subsection{Main sequence stars: chromospheres and impact on planets}
Studies based on spectropolarimetry at visible  -- and more recently near-IR -- wavelengths have allowed us to build a first consistent picture of the dependence of the properties of dynamo-generated magnetic fields on stellar parameters such as mass, rotation period and age \cite{donati09,vidotto14}.
High-resolution UV spectroscopy/spectropolarimetry will enable us to extend this picture to the magnetic and thermodynamic structure of the upper atmosphere -- chromosphere and transition region (TR) -- of cool stars, which is known to be highly structured and dynamic in the solar case thanks to dedicated UV/EUV/X-ray space missions \cite{delZanna18}. 

The UV spectral range indeed contains a number of spectral lines forming at temperatures from $10^4$ to $10^7$~K including chromospheric lines (Mg\textsc{ii}, C\textsc{i}, O\textsc{i}), TR lines (e.g., C\textsc{ii-iv} , N\textsc{iv}, O\textsc{iii-v}, Si\textsc{ii-iv}), the FUV coronal lines of O\textsc{vi} and Fe\textsc{xii}, as well as molecular lines (CO and H$_2$) tracing cool material \cite{pagano04}.
From time-series of high-resolution UV spectra, it is possible to reconstruct the structure of the chromosphere \cite{busa99}. With polarimetry and a full-UV coverage it will be possible to reconstruct the 3D magnetic and thermodynamic structure of upper stellar atmospheres, connect them with spots and magnetic regions observed at the photospheric level with visible spectropolarimetry, and to constrain models of chromospheric and coronal heating \cite{testa15}. In addition, a detailed modelling the astrospheric Lyman $\alpha$ line constitutes the most successful approach so far to measure the mass-loss rate of nearby cool main sequence stars \cite{wood05}. These novel diagnostics will be the basis of new progress on a number of issues:

\begin{itemize}
    \item {\it How do the structure of the upper atmosphere, UV energy output and mass-loss rate of solar-type stars depend on the magnetic topology? How do they change over a range of timescales from rotation period to magnetic cycle and stellar evolution?} Spectropolarimetric surveys carried out at visible wavelengths have provided us with a first picture of the magnetic properties of solar-type stars as a function of their mass and rotation period \cite{marsden14,see15}. Notably, stars with shorter rotation periods than the Sun can exhibit predominantly toroidal topologies, but the effect on the heating of the upper atmosphere, UV flux and mass-loss has not yet been devised \cite{vidotto16}. This connection will have to be explored in the time-domain, with a particular emphasis on stars with well-identified magnetic cycles \cite{boroSaikia18}. While a 1.3~m telescope such as Arago would allow an exploration of the solar neighbourhood, the collecting power of LUVOIR would be key in extending such studies to nearby clusters and thus to build a consistent picture of the joint evolution of rotation, magnetism and UV activity of solar-type stars from the pre-main sequence to the mature main sequence.
    
    \item {\it How do the structure of the upper atmosphere, UV energy output and mass-loss rate of M dwarfs and ultracool dwarfs differ from solar-type stars? How can we reconcile the apparently contradictory activity of ultracool dwarfs as observed at radio and X-ray wavelengths?} The lowest-mass cool stars -- M dwarfs -- constitute an ideal laboratory to test our understanding of stellar magnetism in a very non-solar regime. They are indeed characterized by deep convective envelopes or are even fully-convective, can remain rapid rotators during several Ga, display high-level of activity even at rotation periods of 100~d, and generate radically non-solar magnetic fields \cite{morin08,newton16}. The UV telescope Arago would allow us to target the brightest and most active nearby M dwarfs whose photospheric magnetic fields have already been well-characterized from the ground. With the POLLUX instrument on LUVOIR the study could be extended to a volume-limited sample of nearby M dwarfs currently monitored with the instrument SPIRou \cite{delfosse13}, and to the exploration of the ultracool dwarf regime, for which UV observations could be a key to connect apparently contradictory activity at radio and X-ray wavelengths, with some stars being apparently over-luminous in one of these spectral ranges \cite{williams14}.
    
    \item {\it What is the effect of stellar 
    magnetic activity on the atmospheres of exoplanets? How can we define habitability beyond the insolation criterion?} High-resolution UV spectropolarimetry of cool stars  is also the key to understand how they interact with their planetary systems through stellar winds, flares and associated coronal mass ejections (CMEs), and high energy radiation. These studies will be instrumental in revising the concept of habitable zone beyond the insolation criterion. This issue is particularly acute for M dwarfs whose activity is often cited as a possible obstacle to habitability. Although several studies have addressed the UV emission of these stars and their implications \cite{france18}, we presently lack a consistent picture relating their magnetic properties to their UV emission in the time-domain. For solar-type stars, we still need to describe the evolution of magnetic activity and of its effects on habitability for stars following different rotational tracks from the ZAMS to the mature MS \cite{gallet17}. Finally, ultracool dwarfs have recently appeared as targets of choice to detect rocky planets \cite{gillon16} but presently little is known about their magnetic activity and space weather environment.
\end{itemize}

\subsection{Pre-main sequence stars: star-disk interaction and accretion-ejection}
In addition to probing stellar magnetism in an extremely active regime, high-resolution UV spectropolarimetry of young T~Tauri stars will be a fundamental tool to study the dynamics of accretion shocks and more generally the interaction of the young star with its circumstellar disk. In the magnetospheric accretion model, ionized material is channeled along magnetic field lines onto the stellar surface, and is associated with strong winds and collimated outflows. Gas in the accretion column and in the resulting accretion shock exhibit temperatures in the range $10^4-10^6$~K  emitting a strong blue-ultraviolet continuum along with characteristic emission lines \cite{bouvier14,hartmann16}.

High resolution UV spectroscopy offers a wealth of complementary indicators for the study of T~Tauri stars and their close environment. Their chromosphere and transition region can be probed using the same UV spectral lines as for main sequence stars (see previous section) with Mg~\textsc{ii} h\&k being the most prominent ones. The physical conditions as well as the kinematics of the accretion columns can be determined through the N\textsc{v}, C\textsc{iv}, He\textsc{ii} and Si\textsc{iv} lines, as well as the semi-forbidden lines of C\textsc{ii]}, Fe\textsc{ii]}, and Si\textsc{ii]} in the case of classical T~Tauri stars \cite{lopezMartinez14}. Finally, the Mg\textsc{ii}, Fe\textsc{ii}, and C\textsc{ii} lines as well as the H$_2$ fluorescence lines will allow us to probe the region at the interface between the stellar magnetosphere and the inner rim of the circumstellar disk \cite{najita00}.

\begin{figure*}[t]
    \centering
    \resizebox{\hsize}{!}{\includegraphics{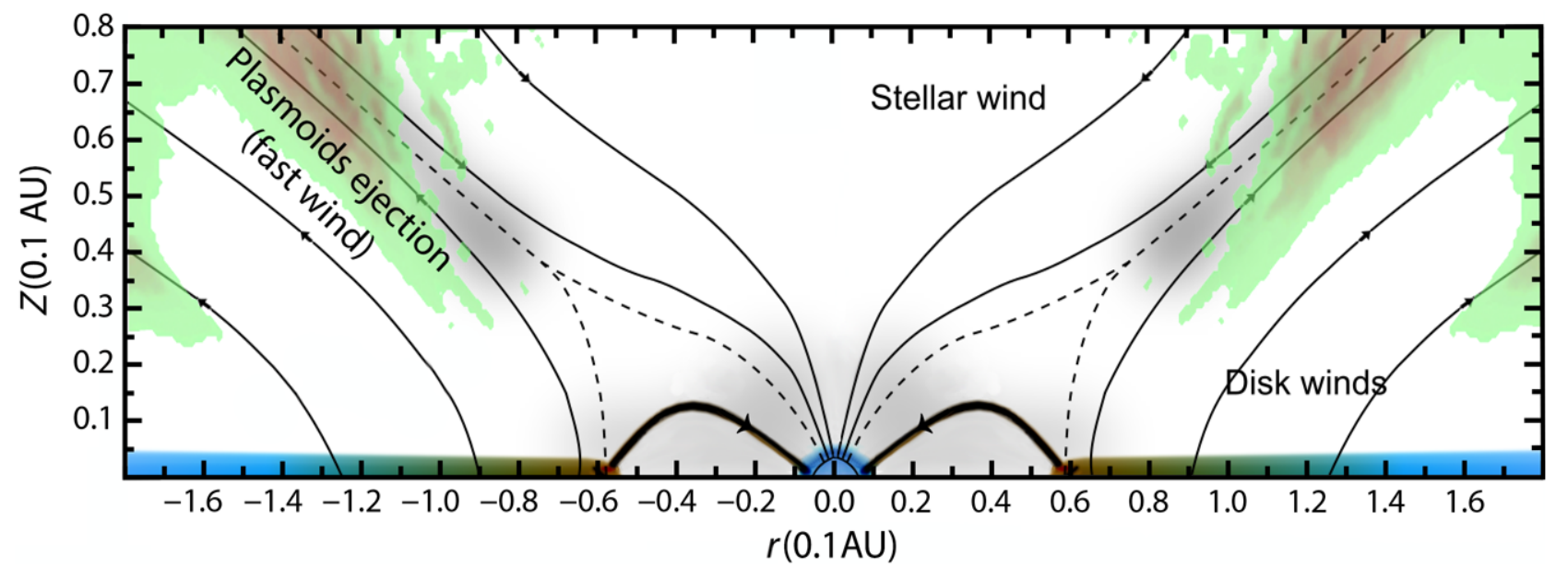}}
    \caption{Schematic view of star-disk interaction and accretion-ejection processes on a T~Tauri star. The color coding represent the emissivity of the C\textsc{iii} line at 191~nm (increasing from light green to brown) as computed from numerical simulations of the interaction between the stellar magnetosphere and the disk, from \cite{gomezDeCastro16}.
}
    \label{fig:tts_accretion}
\end{figure*}

Despite recent progress in both observations and numerical simulations of accretion flows \cite{kurosawa13}, many open questions remain. With time-series of high-resolution UV-visible polarized spectra, it will be possible to study the 3D dynamic structure of accretion shocks and its relation with the surface magnetic field, extending present work in  the visible range \cite{alencar2018}. Such information will be crucial in understanding the relation between stellar parameters, magnetic field properties and the geometry, mass-flux and time-dependence of accretion flows, as well as establishing the physical processes connecting accretion with the wind/outflow process. This point is of prime importance to understand how T~Tauri stars regulate their rotation rate and avoid dislocation while accreting material with high specific angular momentum. Beyond the accretion/ejection processes, such observations will be of prime importance in devising how the high-energy environment of the protostar affects the overall properties of the circumstellar disk and the process of planet formation and migration. More specifically the following issues will be addressed:

\begin{itemize}
    \item {\it How does the magnetic topology affect the magnetospheric accretion-ejection processes?}
    With the first spectropolarimetric surveys of T~Tauri stars carried out with ground-based optical spectropolarimeters we are starting to get a consistent view of the magnetic properties of young Suns as a function of their stellar parameters \cite{gregory12,hill19,villebrun19}, and this effort is presently being pursued at near-Infrared wavelengths \cite{delfosse13}. Combined UV and optical spectropolarimetry will be the ideal tool to establish the relation between the time-dependent 3D structure of the accretion flow and surface magnetic field of the star and test models. With UV spectropolarimetry we will characterize the strong B-fields arising at the sheared interface between star and disk in the PMS phase using emission lines forming within this interface region, and will trace the flows via polarization measurements of the nearby continuum. As stellar rotation decouples from the young planetary disk, the magnetic field is predicted to get stronger and more complex \cite{emeriauViard17}. Spectropolarimetric observations at UV wavelengths  will allow us to investigate the propagation of magnetic energy through the stellar atmosphere into the uppermost coronal layers and the launching of outflows -- winds, magnetospheric ejections and collimated jets \cite{zanni13}.
    
    \item {\it What is the role the stellar magnetic activity in setting the physical conditions in the circumstellar disk and how does this affect the formation and early evolution of planets?} Observing and understanding high-energy radiation originating from the T~Tauri star and its environment is also essential to establish its role in the evolution of the circumstellar disk. This role is twofold: first, high-energy radiation emitted by the star itself and by the collimated jets, as well as energetic particles accelerated during flare events, drive the ionization of the circumstellar disk. This is a key parameter for MHD processes such as the magneto-rotational instability, which is one the main mechanisms invoked to explain the observed mass accretion rates of young T~Tauri stars \cite{bergin03,fraschetti18}. Second, high-energy radiation is thought to be responsible for the evaporation of circumstellar disks \cite{alexander14}. Understanding the connection between between the stellar magnetic field properties and the emission high-energy radiation is therefore crucial to explain the range of timescales observed for disk dissipation from 2 to 10~Ma \cite{salyk09}. Both this disk evolution timescale and the high-energy radiation-driven chemical evolution of the disk are key parameters that impact the formation, early evolution and migration of planets \cite{trilling02, baruteau14}. Novel UV spectropolarimetric observations will constitute a unique opportunity to connect these processes from the stellar magnetic activity to the final architecture of planetary systems.
\end{itemize}
 
\subsection{Evolved stars: chromospheres, surface structures and mass-loss}

\begin{figure*}[t]
    \centering
    \resizebox{0.7\hsize}{!}{\includegraphics{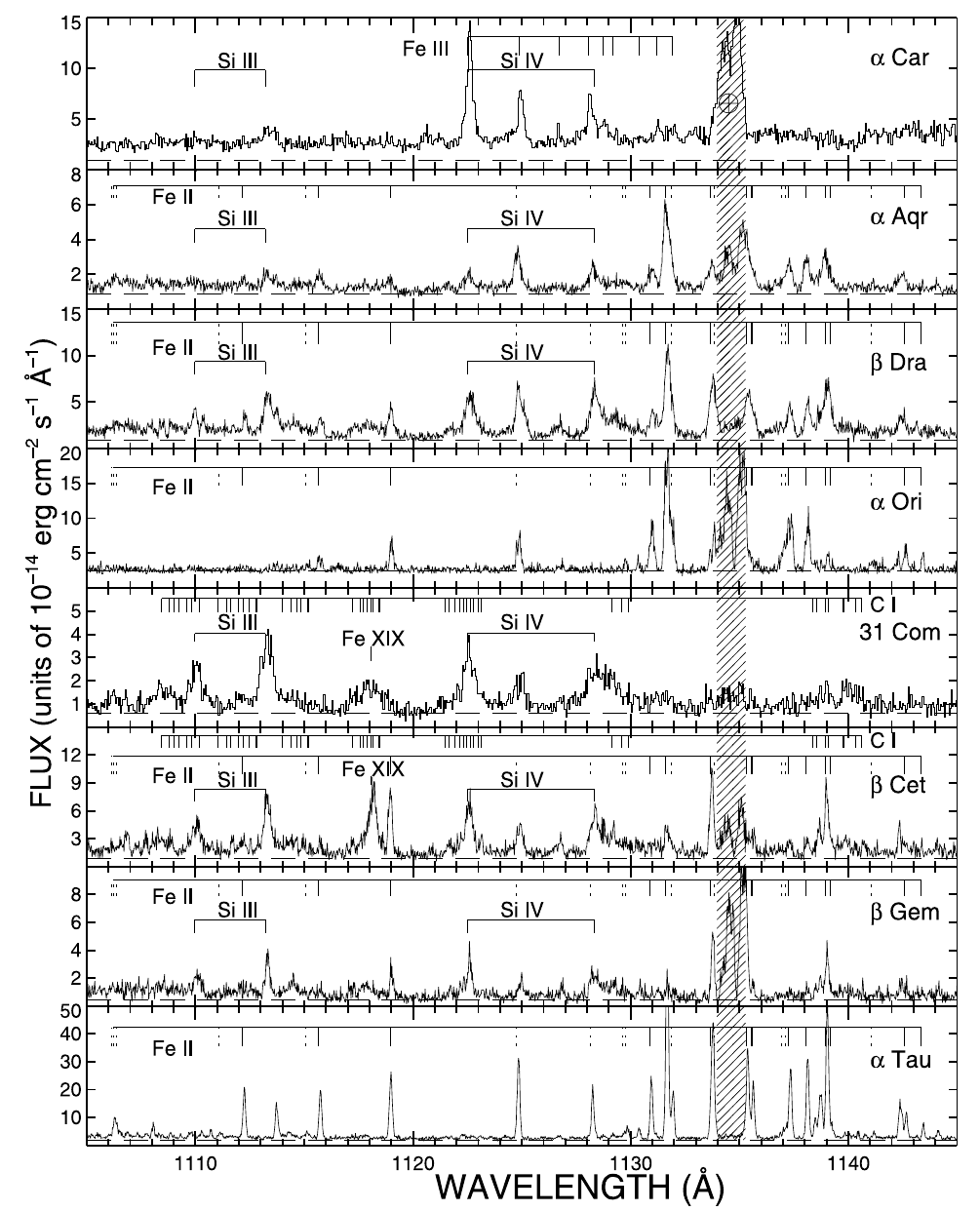}}
    \caption{FUSE (Far Ultraviolet Spectroscopic Explorer) spectra of cool evolved stars in the wavelength region $111-114.5~{\rm nm}$ showing the presence of fluorescent Fe\textsc{ii} emission in many of the coolest targets. The main emission lines are identified and the hatched area marks the position of N\textsc{i} airglow emission, from \cite{dupree05}.
}
    \label{fig:fuse_cool_giants}
\end{figure*}

Cool giant and supergiant stars constitute a late evolutionary stage of low- and high-mass stars respectively. Because of rotational braking during the main sequence (for stars with significant winds during this stage) and angular momentum conservation during the subsequent inflation phase, these stars often display long rotation periods and consequently weak dynamo-generated magnetic fields. As in main sequence stars, with high-resolution UV spectropolarimetry it will be possible to study the 3D dynamic structure of the chromosphere of these stars and address the puzzle of chromospheric heating, but in a regime of parameters very different from solar, with very dynamic atmospheres (shocks and pulsations), generally weak magnetism and active chromospheres often without detectable X-ray coronal counterpart \cite{perezMartinez11}. It would also be possible to investigate the mechanisms -- pulsations, shocks, turbulence and reduced effective gravity, radiation pressure -- that drive the mass-loss of these stars from the photosphere to the upper atmosphere, as well as the role of magnetic fields \cite{josselin07}; while mass-loss would be simultaneously measured through FUV lines \cite{dupree05}. The following issues will be addressed:

\begin{itemize}
    \item {\it How does chromospheric and coronal heating in evolved stars differs from the solar case?} The outer hot atmosphere of single cool evolved stars significantly differs from cool main sequence stars. These slowly rotating objects generate rather weak magnetic fields \cite{konstantinovaAntova14,auriere15} only detectable when their evolutionary track crosses ``magnetic strips'' in the H-R diagram \cite{charbonnel17}, and it is not clear whether solar-like active regions exists at the surface of these stars. Chromospheric heating by acoustic wave dissipation is thought to be an important contributor -- responsible for the ``basal flux'' -- while magnetic processes become important for the most active objects \cite{perezMartinez11}. The detection of coronal FUV lines however points towards the presence of very hot plasma ($\sim 300~{\rm kK}$) in the atmosphere of some of these stars, as does the detection of weak X-ray emission in some objects \cite{dupree05}. Spectroscopic UV observations of cool evolved stars aimed at characterizing the properties of their outer atmospheres as a function of stellar parameters and magnetic field structure are critically needed, since the chromospheric activity of these stars is generally undetectable in the photosphere-dominated optical spectrum. These diagnostics can be efficiently complemented by interferometric observations at submillimetric wavelength which offer the possibility to pinpoint the presence of hot gas in their atmosphere \cite{vlemmings17}.

    \item {\it What are the respective contributions of atmospheric dynamics, radiation pressure and magnetic fields in the mass-loss process of different classes of evolved stars?} Through their mass-loss cool evolved stars are the main contributors to the chemical enrichment of the ISM, and therefore constitute an essential link of the cycle of matter in the Galaxy. Although the main mechanisms likely responsible for this mass-loss have been identified -- involving supersonic convection and shock waves, radiation pressure on molecular lines and dust grains, or magnetic fields -- their respective contribution over the parameter space has not yet been devised \cite{rau19}. With high-resolution spectroscopy at UV wavelengths we will have the unique opportunity to measure the magnetic activity and detect chromospheric plasma, measure atmospheric velocity fields and mass-loss rates, allowing us to study the different waves and pulsation propagation from the photosphere all the way to the upper chromosphere. Besides, the recent discovery of linear polarisation in the spectral lines of cool supergiant and AGB stars -- particularly strong in the blue part of the visible spectrum -- stimulates the development of novel techniques to study the atmospheric dynamics of these stars \cite{lopezAriste18}. Extending such methods to a simultaneous UV-visible coverage appears particularly promising to constrain the link between the large-scale surface dynamics on these stars and mass-loss and to connect them with the structure of their circumstellar environments as observed with high angular resolution techniques \cite{kervella15}.
    
\end{itemize}

\section{Space UV spectropolarimetry}

The planned termination of the highly successful Hubble Space Telescope (HST) mission will eliminate scientific access to large aperture UV optimized space-borne instrumentation. However, there is significant interest in the development of a large UV-optical-IR telescope in the USA, further supported by the US National Academies study "Powering Science" of 2016 emphasized that a multi-purpose “12m class space telescope that would operate from the UV wavelength range to the near-IR domain”, was mandatory to address the most compelling science questions of today Astronomy \& Astrophysics. \\
This concept served as the baseline for the Large Ultraviolet/Optical/Infrared Surveyor (LUVOIR), one of four large mission concepts currently undergoing community study for consideration by the 2020 Astronomy and Astrophysics Decadal Survey, that we will present below. Another, less ambitious, yet powerful, mission concept like CETUS is proposed to NASA, and planned to be operational at a time when there will be no other operational UV capability in the US space science fleet. 

Over the past 40 years, European institutions have worked closely with NASA on developing and observing on UV astrophysics missions such as the International Ultraviolet Explorer (IUE) and the Hubble Space Telescope (HST) in partnership with NASA. There is still keen European interest in UV astrophysics today. For instance, of the 26 UV-related science white papers submitted to NASA's survey Astro2020, about a third were from Europeans. In addition, a couple of mission concept have been developed in Europe such as Arago and EUVO. 

Spectropolarimetry is proposed on all these missions of various sizes and scopes. This technique is a case where Europe is a leader, both concerning the technical development, and science exploitation.

\subsection{POLLUX, a European instrument onboard the LUVOIR NASA flagship project}
\subsubsection{LUVOIR}

LUVOIR is one of four Mission Concept Studies initiated by NASA in January 2016 for its 2020 Decadal Survey of Astronomy and Astrophysics. LUVOIR is designed to be a large multi-wavelength, multi-generational, serviceable observatory following the heritage of the Hubble Space Telescope. In scope with its ambitious planned design, its science goals would enable transformative advances across a broad range of astrophysics. With a proposed launch date in late 2030s, this observatory includes upgradable state-of-the-art instruments and would reside at Earth-Sun L2 point. A large fraction of LUVOIR's schedule would be open to the community through a general observing program. 

The LUVOIR study team is considering two architectures, one with a 15-m mirror (Architecture A), and another with a ~9-m mirror (Architecture B). Architecture A is designed for launch on NASA's planned Space Launch System (SLS), while Architecture B is being designed to launch on a heavy-lift launch vehicle with a 5-m diameter fairing, similar to those in use today. 

    \begin{figure*}[t]
    \centering
    \resizebox{\hsize}{!}{\includegraphics{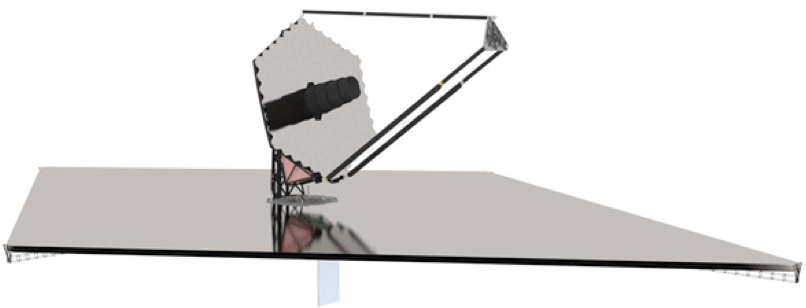}}
    \caption{Preliminary rendering of the LUVOIR Architecture A observatory, which has a 15-meter diameter primary mirror and four instrument bays. An animation of the observatory deployment and pointing may be viewed at \url{https://asd.gsfc.nasa.gov/luvoir/design/}. Credit: A. Jones (NASA GSFC) }
    \label{fig:luvoir}
\end{figure*}

    \begin{figure*}[t]
    \centering
    \resizebox{\hsize}{!}{\includegraphics{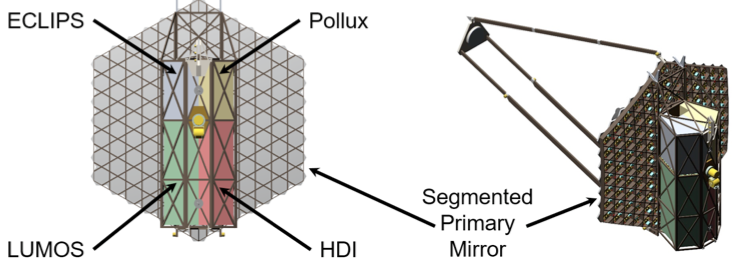}}
    \caption{The four instrument bays of LUVOIR. Major elements of the payload are indicated}
    \label{fig:luvoir_payload}
\end{figure*}

The telescope will be flown in a L2 orbit. A versatile suite of instruments is being developed to fulfill the ambitious scientific goals of this telescope. LUVOIR will allow many scientific breakthroughs in virtually all domains of astrophysics and planetary science, from the epoch of reionization to star and planet formation. One of LUVOIR’s major objectives is to characterize with exquisite details the properties of a large range of exoplanets, including those where life could develop, and be habitable. More precisely, the major objectives of LUVOIR are:  
\begin{itemize}
\item Exploring the full diversity of exoplanets. 
\item Discovering and characterizing exoplanets in the habitable zones of Sun-like stars across a range of ages and searching for bio-signatures in their atmospheres, in a survey large enough to provide evidence for (or against) the presence of habitable planets and life. 
\item Remote sensing of the planets, moons, and minor bodies of the solar system: resolving surface and cloud features as small as 50 km for outer planets and 200~km on Kuiper belt objects, and imaging the icy plumes from giant planet moons.
\item Exploring the building blocks of galaxies both in the local universe and at their emergence in the distant past, and elucidating the nature of dark matter. 
\item Understanding how galaxies form and evolve from active to passive, both by studying their stars and their gaseous fuel across all temperatures and phases. 
\item Following the history of stars in the local volume out to tens of megaparsecs to understand how they form and how they depend on their environment, isolating gravitational wave sources
\item Observing the birth of planets and understanding how the diversity of planetary systems arises. 
\end{itemize}

To reach these goals, LUVOIR will be equipped with instruments offering a total wavelength coverage of $90~{\rm nm} - 2.5~{\rm \mu m}$. Three of these instruments (ECLIPS, HDI, and LUMOS) are presently being studied by NASA while a fourth instrument (POLLUX) is being studied by a European consortium. Those four instruments are:
\begin{itemize}
\item ECLIPS: An ultra-high contrast coronagraph with an imaging camera and integral field spectrograph spanning $200-2,000~{\rm nm}$, capable of directly observing a wide range of exoplanets and obtaining spectra of their atmospheres.
\item HDI: A near-UV to near-IR imager covering $200-2,500~{\rm nm}$, Nyquist sampled at 400~nm and diffraction limited at 500~nm, with high precision astrometry capability. 
\item LUMOS: A far-UV imager and far-UV + near-UV multi-resolution, multi-object spectrograph covering $100-400~{\rm nm}$, capable of simultaneous observations of up to hundreds of sources.
\item POLLUX: A high-resolution UV spectropolarimeter (see details below).
\end{itemize}

The LUVOIR (and the three US-led instruments) study was initiated in January 2016, and was executed by the Goddard Space Flight Center, under the leadership of a Science and Technology Definition Team (STDT) drawn from the community. 
A final report presenting the full study of the telescope and instruments was submitted to NASA HQ by mid-July 2019. See more details at \url{https://wix.to/IkAkAsk}.

We argue that ESA should join NASA in the development of LUVOIR, if selected, since the costs of such a mission will very likely require NASA to seek partners that can make a substantial contribution to the project. We propose that ESA should contribute to the LUVOIR project with and M or L-class contribution, to the benefit the European astronomical community at-large. POLLUX is, to date, the only European-led instrument proposed (and already integrated) to the LUVOIR suite of instruments. 

\subsubsection{POLLUX}
Following discussions between NASA and the French Space Agency (CNES), the POLLUX study was initiated in January 2017. It is supported by CNES (France), and developed by a consortium of European scientists. 

POLLUX is a high-resolution spectropolarimeter operating at UV wavelengths, designed for LUVOIR-A. POLLUX will operate over a broad spectral range (90 to 400~nm), at high spectral resolution (R $\geq$ 120,000). This will allow us to resolve narrow UV emission and absorption lines, enabling us to follow the baryon cycle over cosmic time, from galaxies forming stars out of interstellar gas and grains, and planets forming in circumstellar disks, to the various forms of feedback into the interstellar and intergalactic medium (ISM and IGM), and from active galactic nuclei (AGN). POLLUX will of course also allow us to study many stellar physics issues described above. In particular, since it will be installed on a 15-m telescope, it will be possible to reach stars outside our galaxy. 
  \begin{figure*}[h]
    \centering
    \includegraphics[width=8.cm]{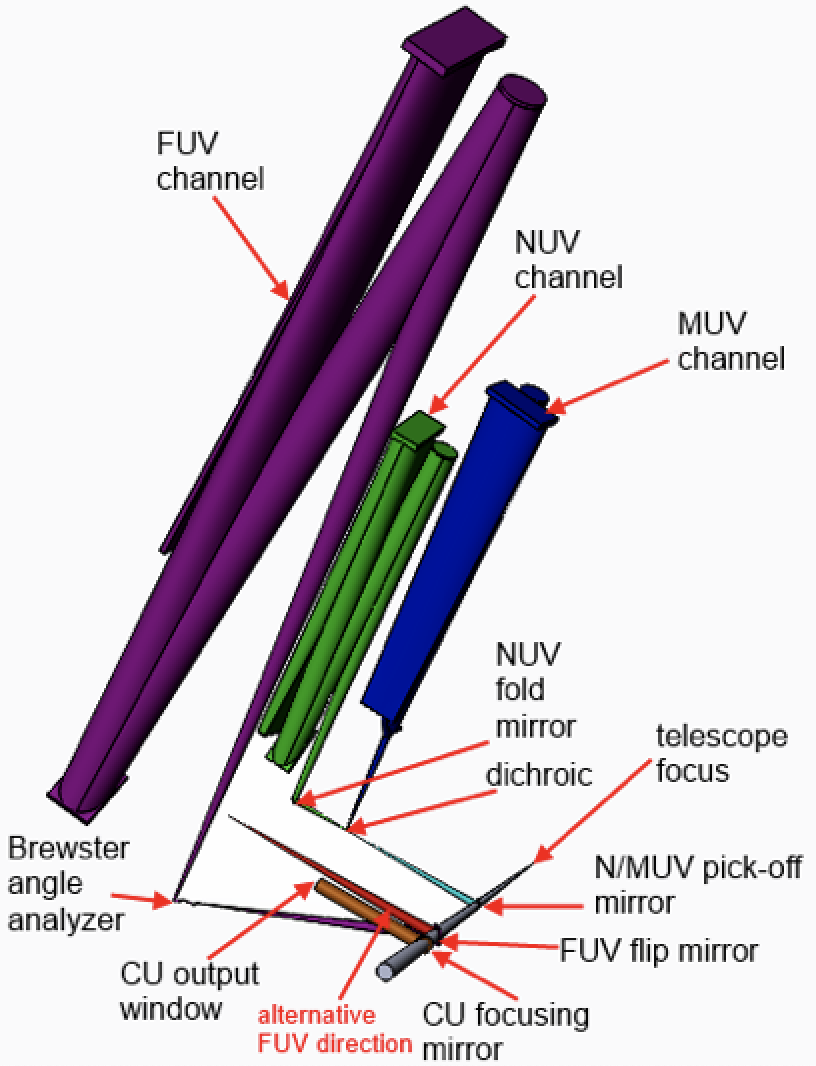}
    \caption{3D rendering of POLLUX optical architecture}
    \label{fig:pollux}
\end{figure*}

The most innovative characteristic of POLLUX is its unique spectropolarimetric capability that will enable the detection of the UV polarized light reflected from exoplanets or from their circumplanetary material, and moons, and characterization of the magnetospheres of stars and planets, and their interactions. UV circular and linear polarization will provide a full picture of magnetic field properties and impact for a variety of media and objects, in particular all types of stars. Linear polarimetry is especially powerful to provide information on deviations from spherical symmetry, providing an extension of interferometry into a domain that is not restricted by the angular size of the objects but by their flux. 
This aspect of POLLUX will be a very powerful tool for studies of the physics and large-scale structure of accretion disks around young stars and white dwarfs, and to constrain the properties of stellar ejecta and explosions. Since the parameter space opened by POLLUX is essentially uncharted territory, its potential for ground-breaking discoveries is tremendous. 

POLLUX is designed for the LUVOIR-A architecture. To define its baseline configuration, we adopted the telescope parameters provided by the LUVOIR study.\\
POLLUX is a spectropolarimeter working in three channels. For practical reasons we refer to these as NUV ($200-400~{\rm nm}$), MUV ($118.5-200~{\rm nm}$), and FUV ($90-124.5~{\rm nm}$). Each channel is equipped with its own dedicated polarimeter followed by a (tailored) high-resolution† spectrograph. This design allows to achieve high spectral resolving power with feasible and affordable values of the detector length, the camera optics field of view, and the overall size of the instrument. It also allows us to use dedicated optical elements, coatings, detector, and polarimeter for each band, hence gaining in efficiency. The MUV + NUV channels are recorded simultaneously, the beams for the two wavelength domains being separated by a dichroic. The dichroic splitter allows the instrument to work in two bands simultaneously and use the full aperture thus achieving the high resolving power with relatively small collimator focal length. 
The FUV spectrum is recorded separately.
POLLUX can be operated in pure spectroscopy mode or in spectropolarimetric mode. The full polarimeters are thus retractable in the MUV and NUV to allow the pure spectroscopic mode. In the FUV only the modulator is retractable. The analyzer is kept in the optical path to direct the beam towards the collimator. 
The full report presenting the POLLUX study is attached to the final LUVOIR report, submitted to NASA in July 2019 for the Decadal review. 
Full details about POLLUX can be also found at \url{https://mission.lam.fr/pollux/}

\subsection{Arago, an M-size mission project for ESA}

Arago is specially designed to address the above science objectives on stars within our galaxy, by obtaining (1) comprehensive 3D maps of selected stars all the way from their sub-photosphere to the frontiers of their immediate circumstellar environment, and (2) high-fidelity multi-parameter information on statistical stellar and planetary samples. Arago's high-resolution spectropolarimeter will be the only facility able to simultaneously deliver all pertinent diagnostics throughout the UV and Visible domains.

\begin{figure*}[h]
    \centering
    \resizebox{0.34\hsize}{!}{\includegraphics[clip]{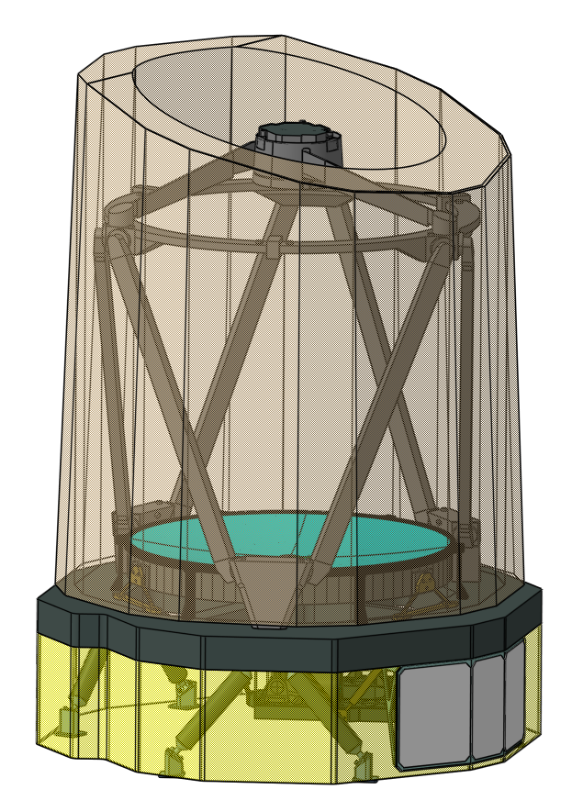}}
    \resizebox{0.63\hsize}{!}{\includegraphics[clip]{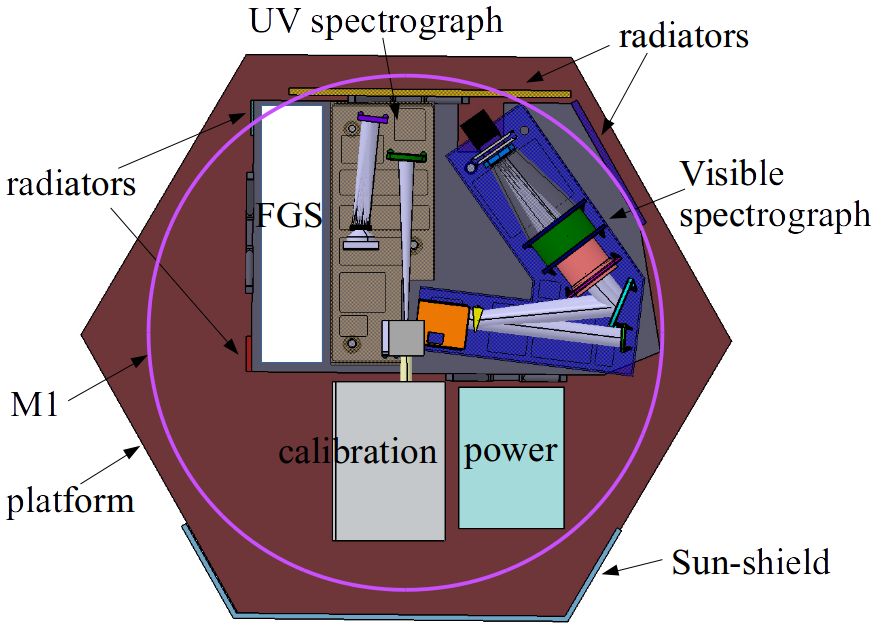}}
    \caption{Left: Arago telescope design, as proposed by ADS. Right: Top view of the instrument layout on the plateform. The size of the primary telescope mirror is also shown.}
    \label{fig:arago_payload}
\end{figure*}

To reach the proposed science goals on galactic stars, it is necessary to observe stellar spectral lines and their polarisation in the Visible and UV wavelength domains simultaneously and with a high-cadence continuous monitoring. The Visible domain allows us to characterise the surface of the star: its properties (e.g., temperature, gravity, rotation, magnetic field) and surface features (e.g., spots, chemical enhancements). The UV domain allows the characterisation of the environment of the star: its wind, magnetosphere, chromosphere, irradiation of exoplanets, etc. Observing both domains simultaneously is the only way to obtain a complete 3D view of the star and its surroundings, and directly link surface features to circumstellar structures, e.g. surface spots to coronal mass ejections, or magnetic footpoints to discs. In addition, the measurement of polarisation in the spectra allows us to detect and quantify the magnetic field and environment of stars and exoplanets. Linear polarimetry provides a means to determine deviations from spherical symmetry of all kinds of objects. It is the extension of interferometry into the domain that is not restricted by the objects’ angular size but merely by its flux.

Arago is a M-class space mission project consisting of a 1.3-meter telescope, equipped with a polarimeter simultaneously feeding two high-resolution spectrographs working in the UV (119-320~nm, R=25,000) and Visible (355-888~nm, R=35,000) spectral ranges. The spacecraft and telescope have been studied by Airbus Defence \& Space (ADS) and Thales Alenia Space (TAS). These industrial studies did not identify any technical issues, and thus confirmed their feasibility (see also Sect~\ref{sec:trl}). The proposed payload consists of a single UV+Visible polarimeter placed near the telescope focal plane. A dichroic behind the polarimeter separates the two wavelength domains to feed two classical echelle spectrometers using cross-dispersion techniques to reach the required spectral resolution. The full spectrum is spread onto two EMCCD detectors. A calibration unit allows us to inject light from calibration lamps in the polarimeter, instead of the stellar light coming from the telescope. The spacecraft and a fine guiding system (FGS) ensures precise pointing stability (30 mas during 30 minutes). The payload has been assessed through a Phase 0 study funded by CNES, which allowed us to refine its design and confirm its feasibility.

The payload, and in particular the detectors, require passive thermal control and thermo-mechanical stability during a complete acquisition sequence (maximum 30 minutes). As the instrument also requires pointing stability and a quiet and stable environment during a complete acquisition sequence, and the scientific objectives require observations anywhere in the sky for up to 30 consecutive days, the proposed mission profile is an L2 Lissajous orbit reached with an Ariane 62 class 1 launcher. 

Arago would mostly observe stars with magnitude between V=3 and 10. It would reach a typical signal-to-noise ratio (SNR) in the intensity spectrum of SNR>100 for B=7 (V=7) mag in 30 minutes for hot (OBA) stars, SNR=100 for B=7 (V=5) mag in 1 hour for solar-like (FG) stars, thus providing a very high SNR for multi-line averaged spectropolarimetric measurements, and SNR=10 in chromospheric emission lines of cool (KM) stars. A magnitude- limited (B$\leq6$ mag) legacy survey would be undertaken and immediately made publicly available. Additional statistical surveys, snapshot targets, targets for detailed 3D mapping, and targets of opportunity (e.g. supernovae) would be chosen following open calls for proposals, in an observatory-type mode. The nominal lifetime of the mission would be 4 years. More information about Arago can be found at \url{http://arago-mission.obspm.fr}.

\subsection{PSS onboard CETUS}

The “Cosmic Evolution Through UV Surveys” (CETUS) is a probe-class mission concept proposed to NASA for its Deacadal 2020 survey with a launch planned in 2029. 

CETUS aims at studying the drivers of galaxy evolution at $z\sim1-2$, make an accurate inventory of baryons in the warm-hot circumgalactic medium (CGM) and find or confirm missing baryons at low redshift, study how the CGM influences galaxy evolution, and measure the light curve of a UV-bright kilonova 200~Mpc away.

CETUS includes a 1.5-m aperture diameter telescope with a large field-of-view. CETUS will provide capabilities for multiple scientific instruments: a Far Ultraviolet (FUV) and Near Ultraviolet (NUV) imaging camera; a NUV Multi-Object Spectrograph (MOS); and a dual-channel Point Source Spectrograph (PSS) in the Lyman Ultraviolet (LUV), FUV, and NUV spectral regions. 
The MOS will take slit spectra of up to 100 sources at once via next-generation Micro-Shutter Array (MSA), over the wavelength range 1800-3500 \AA. Its spectral resolving power will be R $\sim$ 1,000. The wide-field far-UV and near-UV cameras will have a field of view of $17.4'$ x $17.4'$, and will operate over two wavelength ranges, namely 1150-1800, and 2000-4000 \AA. Their angular resolution will be $0.55''$, and $0.33''$, respectively. 
The FUV amd NUV spectrographs can obtain R $\sim$ 20,000 (resp. 40,000) spectra over the wavelength ranges 1000-1800~\AA, and 2000-4000~\AA, and accommodate a $6'$-long slit. 

A collaboration started with France in 2018 with the aim to add a polarimeter to the CETUS NUV spectrograph, similar to the one planned for the POLLUX NUV spectropolarimeter. Discussions are ongoing to move the full PSS UV spectrograph under European leadership.

Each of the three instruments has its own aperture at the telescope focal plane, and each functions independently (with the exception that the prime instrument controls the telescope pointing and roll angle). Each instrument can be removed or inserted into the instrument bay without disturbing the others. Together, the instruments are managed under a single governing ICD and make use of commonality of detectors (CCDs and MCPs), thus having similar electronics, packaging, drivers and software.  Commonalities of Offner relays and devices in the camera and MOS are recognized.

\begin{figure*}[h]
    \centering
    \resizebox{0.35\hsize}{!}{\includegraphics[clip]{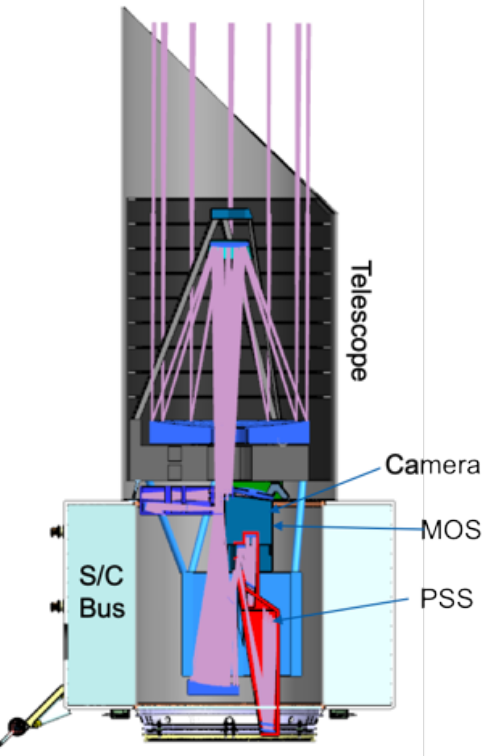}}
    \caption{CETUS telescope design, with its 3 instruments. (Image courtesy by the CETUS team.)}
    \label{fig:cetus_payload}
\end{figure*}

\subsection {EUVO}

The European UltraViolet-Optical Observatory (EUVO) project has been previously submitted to ESA and should be reconsidered for a future L mission. Such an observatory, with a 8-m primary mirror equipped with a high-resolution UV spectropolarimeter, would fulfill many science goals including all the stellar cases described above. It would be ideal to combine both monitoring of many different stellar targets and the possibility to reach faint targets outside our galaxy.

Considering stellar science only, it might however be more cost-effective for ESA to fund POLLUX ($\sim$300 Meuros) on LUVOIR and Arago ($\sim$500 Meuros) than EUVO. Considering the many other science cases that EUVO can address, this L mission nevertheless remains very attractive, especially considering the fact that the selection of LUVOIR by NASA is very uncertain.

\subsection{Readiness and technological challenges}
\label{sec:trl}

The technological solutions required for the different instrument/mission presented above have different levels of technological readiness (TRL). However, solution paths are identified to go from today’s TRL to TRL = 6 by the end of the Preliminary Design Review (PDR) of each of them. 

The LUVOIR and EUVO designs incorporate heritage elements from many missions, including the Hubble Space Telescope (HST), James Webb Space Telescope (JWST), and the Wide Field Infrared Survey Telescope (WFIRST). Use of heritage components establishes confidence in the design’s feasibility, as well as cost and schedule realism. Throughout the concept study, the LUVOIR project has sought to take advantage of heritage designs and components to the greatest extent possible to ensure an appropriate level of risk is constrained to truly new items technology. 

Similarly, the studies by ADS and TAS demonstrated that Arago does not require challenging technology. The spectrographs use classical echelle designs and elements. The UV spectrograph design minimizes the number of optical surfaces to optimize throughput. The UV range requires the use of specific materials (MgF2), coatings, and management of optical surface qualities, but this has already been achieved for other space missions (e.g. IUE, HST).

Except for the microshutter array (NG-MSA), new technologies needed by CETUS are ready now -- from low-scatter gratings to larger and better detectors \cite{Ertley2018}, to telescope mirrors better than Hubble’s, and mirror coatings enabling probes deeper into the far-UV while providing protection against degradation \cite{Fleming2017}. 
The Next-Generation Microshutter Array (MSA) is the key technology for further development that is proceeding at NASA/GSFC with several design improvements and scale-ups already demonstrated and further array fabrication optimizations being addressed in a current 3-year SAT.  Should this not be matured in time, the TRL 8 JWST NIRSpec MSA will be utilized. 

The missions will be equipped with new concepts of spectropolarimeter capable of operating at UV wavelengths, with high spectral resolution power yet delivering high throughputs. The POLLUX instrument is by far the most challenging as it involves the FUV domain down to 90 nm, the technological solutions adopted in the design proposed to NASA in July 2019 have different levels of technological readiness (TRL). 
The most critical technological challenges that have been identified for this high-resolution UV spectropolarimeter are: efficient dichroic and coatings in the UV range, polarimeters, gratings (echelle and cross-dispersers), detectors. 

The mission designs currently rely on electron-multiplication CCDs (EMCCDs) detectors, which are progressively surpassing the capacities of MCPs traditionally used in the UV domain. EMCCDs have an intrinsic gain stage allowing photon counting operation, larger device sizes with smaller pixel dimensions. Recent technological development on entrance window technologies such as $\delta$-doping, have raised their potential competitiveness \cite{nikzad2017}. The technology of $\delta$-doped EMCCDs is not fully mature, and a major challenge will be to demonstrate feasibility of detector wafers large enough to accommodate our needs (typical detector size is 15k x 2k for POLLUX). A promising alternative is to consider CMOS, $\delta$-doped for enhanced UV performance. These devices are rapidly developing and offer larger formats than CCD devices.

The polarimetric technique proposed for all those missions above 123 nm is very similar to polarimeters on-board current space missions (e.g., SOT on JAXA's Hinode or NASA's CLASP), although the instrument may cover a much wider wavelength range. The modulator is composed of a rotating stack of MgF2 plates followed by a polarizing beam-splitter, allowing users to measure the full Stokes (IQUV) spectrum. Below 123~nm however, i.e. for LUVOIR and EUVO, it is necessary to use a reflective polarimeter. Such a reflective UV polarimeter has never flown but is composed of mirrors only, therefore it does not appear particularly challenging. 

\section{Conclusions}
{\bf No space mission equipped with a high-resolution UV spectropolarimeter covering a wide wavelength domain has ever flown. Such an instrument would open a new door in stellar physics by allowing to investigate the origin, evolution, and impact of magnetic fields and magnetospheric structures in both hot and cool stars. This would in particular allow us to take a leap forward in the comprehension of the structure, environment, activity, and evolution of all types of stars, with important consequences on many other domains of astrophysics. A high-resolution UV spectropolarimeter placed on a M-size mission (such as Arago) would allow the detailed 3D mapping of stars and their magnetized environment. Placed on a larger aperture space telescope (such as LUVOIR or a L mission), detailed monitoring would be performed on less stars (as the telescope would likely be shared with other instruments) but it would be possible to reach fainter stars and in particular stars in other galaxies and thus study different ages and metallicities providing additional insights.}

\vfill 

\pagebreak
\section*{Members of the proposing team}

\begin{itemize}
\item Conny Aerts, Institute for Astronomy, KU Leuven, Celestijnlaan 200D, Box 2401, B-3001 Leuven, Belgium
\item Stefano Bagnulo, Armagh Observatory, Northern Ireland, United Kingdom
\item Jean-Claude Bouret, LAM, Aix Marseille University, CNRS, CNES, Marseille, France
\item Claude Catala, LESIA, Paris Observatory, CNRS, PSL University, Sorbonne Universit\'e, Paris University, 5 place Jules Janssen, 92195 Meudon, France
\item Corinne Charbonnel, Geneva Observatory, Switzerland 
\item Chris Evans, UK Astronomy Technology Centre, Royal Observatory, Blackford Hill, Edinburgh, EH9 3HJ, United Kingdom
\item Luca Fossati, Space Research Institute, Austrian Academy of Sciences, Graz, Austria
\item Miriam Garcia, Centro de Astrobiología (CSIC-INTA), Instituto Nacional de Técnica Aeroespacial, 28850 Torrejón de Ardoz (Madrid), Spain
\item Ana I G\'omez de Castro, Space Astronomy Group/AEGORA, Universidad Complutense de Madrid, Spain
\item Artemio Herrero, Instituto de Astrofisica de Canarias \& Universidad de La Laguna, La Laguna, Spain
\item Gaitee Hussain, ESO, Garching, Germany
\item Lex Kaper, Anton Pannekoek Institute for Astronomy, University of Amsterdam, Science Park 904, 1098 XH Amsterdam, The Netherlands
\item Oleg Kochukhov, Department of Physics and Astronomy, Uppsala University, Uppsala, Sweden
\item Renada Konstantinova-Antova, Institute of Astronomy and NAO, BAS, 72
Tsarigradsko shosse blvd., 1784 Sofia, Bulgaria
\item Alex de Koter, Anton Pannekoek Institute for Astronomy, University of Amsterdam, Science Park 904, 1098 XH Amsterdam, The Netherlands, \& Institute for Astronomy, KU Leuven, Celestijnlaan 200D, Box 2401, B-3001 Leuven, Belgium
\item Michaela Kraus, Astronomical Institute, Czech Academy of Sciences, Fri\v{c}ova 298, 251\,65 Ond\v{r}ejov, Czech Republic
\item Ji\v r\'\i\ Krti\v cka, Masaryk University, Kotl\'a\v rsk\'a 2, 611 37, Czech Republic
\item Agnes Lebre, LUPM, Universit\'e de Montpellier, CNRS, 34095 Montpellier, France
\item Theresa Lueftinger, Department of Astrophysics, University of Vienna, Austria
\item Georges Meynet, University of Geneva, Switzerland
\item Julien Morin, LUPM, Universit\'e de Montpellier, CNRS, 34095 Montpellier, France
\item Coralie Neiner, LESIA, Paris Observatory, CNRS, PSL University, Sorbonne Universit\'e, Paris University, 5 place Jules Janssen, 92195 Meudon, France
\item Pascal Petit, IRAP, CNRS, CNES, Universit\'e Toulouse 3, 31400 Toulouse, France
\item Steve Shore, Università di Pisa, Italy
\item Sami Solanki, Max Planck Institute for Solar System Research, Germany
\item Beate Stelzer, Universit\"at T\"ubingen, Germany \& INAF - Osservatorio Astronomico di Palermo, Italy
\item Antoine Strugarek, DRF/IRFU, CEA Paris-Saclay, 91191 Gif-sur-Yvette Cedex, France
\item Aline Vidotto, Trinity College Dublin, Ireland
\item Jorick S Vink, Armagh Observatory \& Planetarium, College Hill, Armagh, Northern Ireland, United Kingdom
\end{itemize}

\pagebreak
\bibliographystyle{unsrt}  
\bibliography{references}  

\begin{thebibliography}{100}

\bibitem{vink01}
J.~S. {Vink}, A.~{de Koter}, and H.~J.~G.~L.~M. {Lamers}.
\newblock {Mass-loss predictions for O and B stars as a function of
  metallicity}.
\newblock {\em \aap}, 369:574--588, April 2001.

\bibitem{langer12}
N.~{Langer}.
\newblock {Presupernova Evolution of Massive Single and Binary Stars}.
\newblock {\em \araa}, 50:107--164, September 2012.

\bibitem{groh13}
Jose~H. {Groh}, Georges {Meynet}, Cyril {Georgy}, and Sylvia {Ekstr{\"o}m}.
\newblock {Fundamental properties of core-collapse supernova and GRB
  progenitors: predicting the look of massive stars before death}.
\newblock {\em \aap}, 558:A131, Oct 2013.

\bibitem{neiner2015}
Coralie {Neiner}, Bram {Buysschaert}, Mary~E. {Oksala}, and Aurore
  {Blaz{\`e}re}.
\newblock {Discovery of two new bright magnetic B stars: i Car and Atlas}.
\newblock {\em \mnras}, 454(1):L56--L60, Nov 2015.

\bibitem{grunhut15}
Jason~H. {Grunhut} and Coralie {Neiner}.
\newblock {Magnetic fields in early-type stars}.
\newblock In K.~N. {Nagendra}, Stefano {Bagnulo}, Rebecca {Centeno}, and
  Mar{\'\i}a. {Jes{\'u}s Mart{\'\i}nez Gonz{\'a}lez}, editors, {\em
  Polarimetry}, volume 305 of {\em IAU Symposium}, pages 53--60, Oct 2015.

\bibitem{blazere2016}
A.~{Blaz{\`e}re}, P.~{Petit}, F.~{Ligni{\`e}res}, M.~{Auri{\`e}re},
  J.~{Ballot}, T.~{B{\"o}hm}, C.~P. {Folsom}, M.~{Gaurat}, L.~{Jouve},
  A.~{Lopez Ariste}, C.~{Neiner}, and G.~A. {Wade}.
\newblock {Detection of ultra-weak magnetic fields in Am stars:
  {\ensuremath{\beta}} Ursae Majoris and {\ensuremath{\theta}} Leonis}.
\newblock {\em \aap}, 586:A97, Feb 2016.

\bibitem{maeder03}
A.~{Maeder} and G.~{Meynet}.
\newblock {Stellar evolution with rotation and magnetic fields. I. The relative
  importance of rotational and magnetic effects}.
\newblock {\em \aap}, 411:543--552, Dec 2003.

\bibitem{song16}
H.~F. {Song}, G.~{Meynet}, A.~{Maeder}, S.~{Ekstr{\"o}m}, and P.~{Eggenberger}.
\newblock {Massive star evolution in close binaries. Conditions for homogeneous
  chemical evolution}.
\newblock {\em \aap}, 585:A120, Jan 2016.

\bibitem{mestel99}
Leon {Mestel}.
\newblock {\em {Stellar magnetism}}.
\newblock Oxford University Press, Oxford, 1999.

\bibitem{heger05}
A.~{Heger}, S.~E. {Woosley}, and H.~C. {Spruit}.
\newblock {Presupernova Evolution of Differentially Rotating Massive Stars
  Including Magnetic Fields}.
\newblock {\em \apj}, 626(1):350--363, Jun 2005.

\bibitem{mathis05}
S.~{Mathis} and J.~P. {Zahn}.
\newblock {Transport and mixing in the radiation zones of rotating stars. II.
  Axisymmetric magnetic field}.
\newblock {\em \aap}, 440(2):653--666, Sep 2005.

\bibitem{uddoula02}
Asif {ud-Doula} and Stanley~P. {Owocki}.
\newblock {Dynamical Simulations of Magnetically Channeled Line-driven Stellar
  Winds. I. Isothermal, Nonrotating, Radially Driven Flow}.
\newblock {\em \apj}, 576(1):413--428, Sep 2002.

\bibitem{uddoula08}
Asif {Ud-Doula}, Stanley~P. {Owocki}, and Richard H.~D. {Townsend}.
\newblock {Dynamical simulations of magnetically channelled line-driven stellar
  winds - II. The effects of field-aligned rotation}.
\newblock {\em \mnras}, 385(1):97--108, Mar 2008.

\bibitem{uddoula09}
Asif {Ud-Doula}, Stanley~P. {Owocki}, and Richard H.~D. {Townsend}.
\newblock {Dynamical simulations of magnetically channelled line-driven stellar
  winds - III. Angular momentum loss and rotational spin-down}.
\newblock {\em \mnras}, 392(3):1022--1033, Jan 2009.

\bibitem{meynet11}
G.~{Meynet}, P.~{Eggenberger}, and A.~{Maeder}.
\newblock {Massive star models with magnetic braking}.
\newblock {\em \aap}, 525:L11, Jan 2011.

\bibitem{townsend10}
R.~H.~D. {Townsend}, M.~E. {Oksala}, D.~H. {Cohen}, S.~P. {Owocki}, and
  A.~{ud-Doula}.
\newblock {Discovery of Rotational Braking in the Magnetic Helium-strong Star
  Sigma Orionis E}.
\newblock {\em \apjl}, 714(2):L318--L322, May 2010.

\bibitem{petermann15}
I.~{Petermann}, N.~{Langer}, N.~{Castro}, and L.~{Fossati}.
\newblock {Blue supergiants as descendants of magnetic main sequence stars}.
\newblock {\em \aap}, 584:A54, Dec 2015.

\bibitem{maeder14}
Andr{\'e} {Maeder} and Georges {Meynet}.
\newblock {Magnetic Braking of Stellar Cores in Red Giants and Supergiants}.
\newblock {\em \apj}, 793(2):123, Oct 2014.

\bibitem{alecian15}
E.~{Alecian}, C.~{Neiner}, G.~A. {Wade}, S.~{Mathis}, D.~{Bohlender},
  D.~{C{\'e}bron}, C.~{Folsom}, J.~{Grunhut}, J.~B. {Le Bouquin}, V.~{Petit},
  H.~{Sana}, A.~{Tkachenko}, and A.~{ud-Doula}.
\newblock {The BinaMIcS project: understanding the origin of magnetic fields in
  massive stars through close binary systems}.
\newblock In Georges {Meynet}, Cyril {Georgy}, Jos{\'e} {Groh}, and Philippe
  {Stee}, editors, {\em New Windows on Massive Stars}, volume 307 of {\em IAU
  Symposium}, pages 330--335, Jan 2015.

\bibitem{commercon11}
Beno{\^\i}t {Commer{\c{c}}on}, Patrick {Hennebelle}, and Thomas {Henning}.
\newblock {Collapse of Massive Magnetized Dense Cores Using Radiation
  Magnetohydrodynamics: Early Fragmentation Inhibition}.
\newblock {\em \apjl}, 742(1):L9, Nov 2011.

\bibitem{blazere15}
A.~{Blaz{\`e}re}, C.~{Neiner}, A.~{Tkachenko}, J.~C. {Bouret}, and Th.
  {Rivinius}.
\newblock {The magnetic field of {\ensuremath{\zeta}} Orionis A}.
\newblock {\em \aap}, 582:A110, Oct 2015.

\bibitem{fossati16}
L.~{Fossati}, F.~R.~N. {Schneider}, N.~{Castro}, N.~{Langer},
  S.~{Sim{\'o}n-D{\'\i}az}, A.~{M{\"u}ller}, A.~{de Koter}, T.~{Morel},
  V.~{Petit}, H.~{Sana}, and G.~A. {Wade}.
\newblock {Evidence of magnetic field decay in massive main-sequence stars}.
\newblock {\em \aap}, 592:A84, Aug 2016.

\bibitem{shultz19}
M.~{Shultz}, J.~B. {Le Bouquin}, Th~{Rivinius}, G.~A. {Wade}, O.~{Kochukhov},
  E.~{Alecian}, V.~{Petit}, O.~{Pfuhl}, M.~{Karl}, F.~{Gao}, R.~{Grellmann},
  C.~C. {Lin}, P.~{Garcia}, S.~{Lacour}, {MiMeS Collaboration}, and {BinaMIcS
  Collaboration}.
\newblock {NU Ori: a hierarchical triple system with a strongly magnetic B-type
  star}.
\newblock {\em \mnras}, 482(3):3950--3965, Jan 2019.

\bibitem{featherstone09}
Nicholas~A. {Featherstone}, Matthew~K. {Browning}, Allan~Sacha {Brun}, and Juri
  {Toomre}.
\newblock {Effects of Fossil Magnetic Fields on Convective Core Dynamos in
  A-type Stars}.
\newblock {\em \apj}, 705(1):1000--1018, Nov 2009.

\bibitem{auriere08}
M.~{Auri{\`e}re}, G.~A. {Wade}, F.~{Ligni{\`e}res}, J.~D. {Land street}, J.~F.
  {Donati}, A.~{Hui Bon Hoa}, I.~{Iliev}, P.~{Petit}, T.~{Roudier},
  J.~{Silvester}, and S.~{Theado}.
\newblock {Weak magnetic fields in CP stars}.
\newblock {\em Contributions of the Astronomical Observatory Skalnate Pleso},
  38(2):211--216, Apr 2008.

\bibitem{neiner18}
C.~{Neiner}, A.~{Martin}, G.~{Wade}, and M.~{Oksala}.
\newblock {The magnetic field of evolved hot stars}.
\newblock In {\em SF2A-2018: Proceedings of the Annual meeting of the French
  Society of Astronomy and Astrophysics}, page~Di, Dec 2018.

\bibitem{keszthelyi19}
Z.~{Keszthelyi}, G.~{Meynet}, C.~{Georgy}, G.~A. {Wade}, V.~{Petit}, and
  A.~{David-Uraz}.
\newblock {The effects of surface fossil magnetic fields on massive star
  evolution: I. Magnetic field evolution, mass-loss quenching, and magnetic
  braking}.
\newblock {\em \mnras}, 485(4):5843--5860, Jun 2019.

\bibitem{martins13}
F.~{Martins} and A.~{Palacios}.
\newblock {A comparison of evolutionary tracks for single Galactic massive
  stars}.
\newblock {\em \aap}, 560:A16, December 2013.

\bibitem{bard16}
Christopher {Bard} and Richard H.~D. {Townsend}.
\newblock {Effect of a magnetic field on massive-star winds - I. Mass-loss and
  velocity for a dipole field}.
\newblock {\em \mnras}, 462(4):3672--3688, Nov 2016.

\bibitem{uddoula13}
A.~{ud-Doula}, J.~O. {Sundqvist}, S.~P. {Owocki}, V.~{Petit}, and R.~H.~D.
  {Townsend}.
\newblock {First 3DMHD simulation of a massive-star magnetosphere with
  application to H{\ensuremath{\alpha}} emission from
  {\ensuremath{\theta}}$^{1}$ Ori C}.
\newblock {\em \mnras}, 428(3):2723--2730, Jan 2013.

\bibitem{petit17}
V.~{Petit}, Z.~{Keszthelyi}, R.~{MacInnis}, D.~H. {Cohen}, R.~H.~D. {Townsend},
  G.~A. {Wade}, S.~L. {Thomas}, S.~P. {Owocki}, J.~{Puls}, and A.~{ud-Doula}.
\newblock {Magnetic massive stars as progenitors of `heavy' stellar-mass black
  holes}.
\newblock {\em \mnras}, 466(1):1052--1060, Apr 2017.

\bibitem{georgy17}
Cyril {Georgy}, Georges {Meynet}, Sylvia {Ekstr{\"o}m}, Gregg~A. {Wade},
  V{\'e}ronique {Petit}, Zsolt {Keszthelyi}, and Raphael {Hirschi}.
\newblock {Possible pair-instability supernovae at solar metallicity from
  magnetic stellar progenitors}.
\newblock {\em \aap}, 599:L5, Mar 2017.

\bibitem{abbott16}
B.~P. Abbott, R.~Abbott, T.~D. Abbott, M.~R. Abernathy, {LIGO Scientific
  Collaboration}, and {Virgo Collaboration}.
\newblock Observation of gravitational waves from a binary black hole merger.
\newblock {\em \prl}, 116(6):061102, Feb 2016.

\bibitem{abbott17a}
B.~P. {Abbott}, R.~{Abbott}, T.~D. {Abbott}, F.~{Acernese}, K.~{Ackley},
  C.~{Adams}, T.~{Adams}, P.~{Addesso}, {LIGO Scientific Collaboration}, and
  {Virgo Collaboration}.
\newblock {GW170817: Observation of Gravitational Waves from a Binary Neutron
  Star Inspiral}.
\newblock {\em \prl}, 119(16):161101, Oct 2017.

\bibitem{abbott17b}
B.~P. {Abbott}, R.~{Abbott}, T.~D. {Abbott}, F.~{Acernese}, K.~{Ackley},
  C.~{Adams}, T.~{Adams}, P.~{Addesso}, {(LIGO Scientific Collaboration}, and
  {Virgo Collaboration}.
\newblock {GW170608: Observation of a 19 Solar-mass Binary Black Hole
  Coalescence}.
\newblock {\em \apjl}, 851(2):L35, Dec 2017.

\bibitem{ferrario15}
Lilia {Ferrario}, Domitilla {de Martino}, and Boris~T. {G{\"a}nsicke}.
\newblock {Magnetic White Dwarfs}.
\newblock {\em \ssr}, 191(1-4):111--169, Oct 2015.

\bibitem{wickramasinghe05}
D.~T. {Wickramasinghe} and Lilia {Ferrario}.
\newblock {The origin of the magnetic fields in white dwarfs}.
\newblock {\em \mnras}, 356(4):1576--1582, Feb 2005.

\bibitem{tout08}
C.~A. {Tout}, D.~T. {Wickramasinghe}, J.~{Liebert}, L.~{Ferrario}, and J.~E.
  {Pringle}.
\newblock {Binary star origin of high field magnetic white dwarfs}.
\newblock {\em \mnras}, 387(2):897--901, Jun 2008.

\bibitem{briggs15}
Gordon~P. {Briggs}, Lilia {Ferrario}, Christopher~A. {Tout}, Dayal~T.
  {Wickramasinghe}, and Jarrod~R. {Hurley}.
\newblock {Merging binary stars and the magnetic white dwarfs}.
\newblock {\em \mnras}, 447(2):1713--1723, Feb 2015.

\bibitem{valyavin14}
G.~{Valyavin}, D.~{Shulyak}, G.~A. {Wade}, K.~{Antonyuk}, S.~V. {Zharikov},
  G.~A. {Galazutdinov}, S.~{Plachinda}, S.~{Bagnulo}, L.~{Fox Machado},
  M.~{Alvarez}, D.~M. {Clark}, J.~M. {Lopez}, D.~{Hiriart}, Inwoo {Han},
  Young-Beom {Jeon}, C.~{Zurita}, R.~{Mujica}, T.~{Burlakova}, T.~{Szeifert},
  and A.~{Burenkov}.
\newblock {Suppression of cooling by strong magnetic fields in white dwarf
  stars}.
\newblock {\em \nat}, 515(7525):88--91, Nov 2014.

\bibitem{guilet10}
J{\'e}r{\^o}me {Guilet}, Jun'ichi {Sato}, and Thierry {Foglizzo}.
\newblock {The Saturation of SASI by Parasitic Instabilities}.
\newblock {\em \apj}, 713(2):1350--1362, Apr 2010.

\bibitem{burrows07}
A.~{Burrows}, L.~{Dessart}, E.~{Livne}, C.~D. {Ott}, and J.~{Murphy}.
\newblock {Simulations of Magnetically Driven Supernova and Hypernova
  Explosions in the Context of Rapid Rotation}.
\newblock {\em \apj}, 664(1):416--434, Jul 2007.

\bibitem{mosta15}
Philipp {M{\"o}sta}, Christian~D. {Ott}, David {Radice}, Luke~F. {Roberts},
  Erik {Schnetter}, and Roland {Haas}.
\newblock {A large-scale dynamo and magnetoturbulence in rapidly rotating
  core-collapse supernovae}.
\newblock {\em \nat}, 528(7582):376--379, Dec 2015.

\bibitem{couch13}
Sean~M. {Couch} and Christian~D. {Ott}.
\newblock {Revival of the Stalled Core-collapse Supernova Shock Triggered by
  Precollapse Asphericity in the Progenitor Star}.
\newblock {\em \apjl}, 778(1):L7, Nov 2013.

\bibitem{couch15}
Sean~M. {Couch}, Emmanouil {Chatzopoulos}, W.~David {Arnett}, and F.~X.
  {Timmes}.
\newblock {The Three-dimensional Evolution to Core Collapse of a Massive Star}.
\newblock {\em \apjl}, 808(1):L21, Jul 2015.

\bibitem{mueller15}
B.~{M{\"u}ller} and H.~Th. {Janka}.
\newblock {Non-radial instabilities and progenitor asphericities in
  core-collapse supernovae}.
\newblock {\em \mnras}, 448(3):2141--2174, Apr 2015.

\bibitem{blondin07}
John~M. {Blondin} and Anthony {Mezzacappa}.
\newblock {Pulsar spins from an instability in the accretion shock of
  supernovae}.
\newblock {\em \nat}, 445(7123):58--60, Jan 2007.

\bibitem{yamasaki08}
Tatsuya {Yamasaki} and Thierry {Foglizzo}.
\newblock {Effect of Rotation on the Stability of a Stalled Cylindrical Shock
  and Its Consequences for Core-Collapse Supernovae}.
\newblock {\em \apj}, 679(1):607--615, May 2008.

\bibitem{iwakami14}
Wakana {Iwakami}, Hiroki {Nagakura}, and Shoichi {Yamada}.
\newblock {Parametric Study of Flow Patterns behind the Standing Accretion
  Shock Wave for Core-Collapse Supernovae}.
\newblock {\em \apj}, 786(2):118, May 2014.

\bibitem{nakamura14}
Satoshi~X. {Nakamura}.
\newblock {Neutrino Emissivities from Deuteron-Breakup and Formation in
  Supernovae}.
\newblock In {\em Journal of Physics Conference Series}, volume 569, page
  012057, Dec 2014.

\bibitem{takiwaki16}
Tomoya {Takiwaki}, Kei {Kotake}, and Yudai {Suwa}.
\newblock {Three-dimensional simulations of rapidly rotating core-collapse
  supernovae: finding a neutrino-powered explosion aided by non-axisymmetric
  flows}.
\newblock {\em \mnras}, 461(1):L112--L116, Sep 2016.

\bibitem{asplund04}
M.~{Asplund}, N.~{Grevesse}, A.~J. {Sauval}, C.~{Allende Prieto}, and
  D.~{Kiselman}.
\newblock {Line formation in solar granulation. IV. [O I], O I and OH lines and
  the photospheric O abundance}.
\newblock {\em \aap}, 417:751--768, Apr 2004.

\bibitem{mestel05}
Leon Mestel and John D.~Landstreet.
\newblock {\em Stellar Magnetic Fields}, pages 183--218.
\newblock Springer Berlin Heidelberg, Berlin, Heidelberg, 2005.

\bibitem{donati09}
J.~F. {Donati} and J.~D. {Landstreet}.
\newblock {Magnetic Fields of Nondegenerate Stars}.
\newblock {\em \araa}, 47(1):333--370, Sep 2009.

\bibitem{vidotto14}
A.~A. {Vidotto}, S.~G. {Gregory}, M.~{Jardine}, J.~F. {Donati}, P.~{Petit},
  J.~{Morin}, C.~P. {Folsom}, J.~{Bouvier}, A.~C. {Cameron}, G.~{Hussain},
  S.~{Marsden}, I.~A. {Waite}, R.~{Fares}, S.~{Jeffers}, and J.~D. {do
  Nascimento}.
\newblock {Stellar magnetism: empirical trends with age and rotation}.
\newblock {\em \mnras}, 441(3):2361--2374, Jul 2014.

\bibitem{delZanna18}
Giulio {Del Zanna} and Helen~E. {Mason}.
\newblock {Solar UV and X-ray spectral diagnostics}.
\newblock {\em Living Reviews in Solar Physics}, 15(1):5, Aug 2018.

\bibitem{pagano04}
I.~{Pagano}, J.~L. {Linsky}, J.~{Valenti}, and D.~K. {Duncan}.
\newblock {HST/STIS high resolution echelle spectra of
  <ASTROBJ>{\ensuremath{\alpha}} Centauri A</ASTROBJ> (G2 V)}.
\newblock {\em \aap}, 415:331--348, Feb 2004.

\bibitem{busa99}
I.~{Bus{\`a}}, I.~{Pagano}, M.~{Rodon{\`o}}, J.~E. {Neff}, and A.~C.
  {Lanzafame}.
\newblock {Chromospheric imaging of the active binary system V 711 Tauri = HR
  1099 in December 1992}.
\newblock {\em \aap}, 350:571--581, Oct 1999.

\bibitem{testa15}
P.~{Testa}, B.~{De Pontieu}, and V.~H. {Hansteen}.
\newblock {Properties of moss emission from joint FeXII IRIS and Hinode
  observations of active region plasma}.
\newblock In {\em AGU Fall Meeting Abstracts}, volume 2015, pages SH31D--06,
  Dec 2015.

\bibitem{wood05}
B.~E. {Wood}, H.~R. {M{\"u}ller}, G.~P. {Zank}, J.~L. {Linsky}, and
  S.~{Redfield}.
\newblock {New Mass-Loss Measurements from Astrospheric Ly{\ensuremath{\alpha}}
  Absorption}.
\newblock {\em \apjl}, 628(2):L143--L146, Aug 2005.

\bibitem{marsden14}
S.~C. {Marsden}, P.~{Petit}, S.~V. {Jeffers}, J.~{Morin}, R.~{Fares},
  A.~{Reiners}, J.~D. {do Nascimento}, M.~{Auri{\`e}re}, J.~{Bouvier}, B.~D.
  {Carter}, C.~{Catala}, B.~{Dintrans}, J.~F. {Donati}, T.~{Gastine},
  M.~{Jardine}, R.~{Konstantinova-Antova}, J.~{Lanoux}, F.~{Ligni{\`e}res},
  A.~{Morgenthaler}, J.~C. {Ram{\`\i}rez-V{\`e}lez}, S.~{Th{\'e}ado}, V.~{Van
  Grootel}, and {BCool Collaboration}.
\newblock {A BCool magnetic snapshot survey of solar-type stars}.
\newblock {\em \mnras}, 444(4):3517--3536, Nov 2014.

\bibitem{see15}
V.~{See}, M.~{Jardine}, A.~A. {Vidotto}, J.-F. {Donati}, C.~P. {Folsom},
  S.~{Boro Saikia}, J.~{Bouvier}, R.~{Fares}, S.~G. {Gregory}, G.~{Hussain},
  S.~V. {Jeffers}, S.~C. {Marsden}, J.~{Morin}, C.~{Moutou}, J.~D. {do
  Nascimento}, P.~{Petit}, L.~{Ros{\'e}n}, and I.~A. {Waite}.
\newblock {The energy budget of stellar magnetic fields}.
\newblock {\em \mnras}, 453:4301--4310, November 2015.

\bibitem{vidotto16}
A.~A. {Vidotto}, J.~F. {Donati}, M.~{Jardine}, V.~{See}, P.~{Petit},
  I.~{Boisse}, S.~{Boro Saikia}, E.~{H{\'e}brard}, S.~V. {Jeffers}, S.~C.
  {Marsden}, and J.~{Morin}.
\newblock {Could a change in magnetic field geometry cause the break in the
  wind-activity relation?}
\newblock {\em \mnras}, 455(1):L52--L56, Jan 2016.

\bibitem{boroSaikia18}
S.~{Boro Saikia}, T.~{Lueftinger}, S.~V. {Jeffers}, C.~P. {Folsom}, V.~{See},
  P.~{Petit}, S.~C. {Marsden}, A.~A. {Vidotto}, J.~{Morin}, A.~{Reiners},
  M.~{Guedel}, and {BCool Collaboration}.
\newblock {Direct evidence of a full dipole flip during the magnetic cycle of a
  sun-like star}.
\newblock {\em \aap}, 620:L11, Dec 2018.

\bibitem{morin08}
J.~{Morin}, J.~F. {Donati}, P.~{Petit}, X.~{Delfosse}, T.~{Forveille},
  L.~{Albert}, M.~{Auri{\`e}re}, R.~{Cabanac}, B.~{Dintrans}, R.~{Fares},
  T.~{Gastine}, M.~M. {Jardine}, F.~{Ligni{\`e}res}, F.~{Paletou}, J.~C.
  {Ramirez Velez}, and S.~{Th{\'e}ado}.
\newblock {Large-scale magnetic topologies of mid M dwarfs}.
\newblock {\em \mnras}, 390(2):567--581, Oct 2008.

\bibitem{newton16}
Elisabeth~R. {Newton}, Jonathan {Irwin}, David {Charbonneau}, Zachory~K.
  {Berta-Thompson}, Jason~A. {Dittmann}, and Andrew~A. {West}.
\newblock {The Rotation and Galactic Kinematics of Mid M Dwarfs in the Solar
  Neighborhood}.
\newblock {\em \apj}, 821(2):93, Apr 2016.

\bibitem{delfosse13}
X.~{Delfosse}, J.~F. {Donati}, D.~{Kouach}, G.~{H{\'e}brard}, R.~{Doyon},
  E.~{Artigau}, F.~{Bouchy}, I.~{Boisse}, A.~S. {Brun}, P.~{Hennebelle},
  T.~{Widemann}, J.~{Bouvier}, X.~{Bonfils}, J.~{Morin}, C.~{Moutou},
  F.~{Pepe}, S.~{Udry}, J.~D. {do Nascimento}, S.~H.~P. {Alencar}, B.~V.
  {Castilho}, E.~{Martioli}, S.~Y. {Wang}, P.~{Figueira}, and N.~C. {Santos}.
\newblock {World-leading science with SPIRou - The nIR spectropolarimeter /
  high-precision velocimeter for CFHT}.
\newblock In L.~{Cambresy}, F.~{Martins}, E.~{Nuss}, and A.~{Palacios},
  editors, {\em SF2A-2013: Proceedings of the Annual meeting of the French
  Society of Astronomy and Astrophysics}, pages 497--508, Nov 2013.

\bibitem{williams14}
P.~K.~G. {Williams}, B.~A. {Cook}, and E.~{Berger}.
\newblock {Trends in Ultracool Dwarf Magnetism. I. X-Ray Suppression and Radio
  Enhancement}.
\newblock {\em \apj}, 785(1):9, Apr 2014.

\bibitem{france18}
Kevin {France}, Nicole {Arulanantham}, Luca {Fossati}, Antonino~F. {Lanza},
  R.~O.~Parke {Loyd}, Seth {Redfield}, and P.~Christian {Schneider}.
\newblock {Far-ultraviolet Activity Levels of F, G, K, and M Dwarf Exoplanet
  Host Stars}.
\newblock {\em \apjs}, 239(1):16, Nov 2018.

\bibitem{gallet17}
F.~{Gallet}, C.~{Charbonnel}, L.~{Amard}, S.~{Brun}, A.~{Palacios}, and
  S.~{Mathis}.
\newblock {Impacts of stellar evolution and dynamics on the habitable zone: The
  role of rotation and magnetic activity}.
\newblock {\em \aap}, 597:A14, Jan 2017.

\bibitem{gillon16}
Micha{\"e}l {Gillon}, Emmanu{\"e}l {Jehin}, Susan~M. {Lederer}, Laetitia
  {Delrez}, Julien {de Wit}, Artem {Burdanov}, Val{\'e}rie {Van Grootel},
  Adam~J. {Burgasser}, Amaury H.~M.~J. {Triaud}, Cyrielle {Opitom},
  Brice-Olivier {Demory}, Devendra~K. {Sahu}, Daniella {Bardalez Gagliuffi},
  Pierre {Magain}, and Didier {Queloz}.
\newblock {Temperate Earth-sized planets transiting a nearby ultracool dwarf
  star}.
\newblock {\em \nat}, 533(7602):221--224, May 2016.

\bibitem{bouvier14}
J.~{Bouvier}.
\newblock {The magnetospheric accretion/ejection process in young stellar
  objects: open issues and perspectives}.
\newblock In {\em European Physical Journal Web of Conferences}, volume~64 of
  {\em European Physical Journal Web of Conferences}, page 09001, Jan 2014.

\bibitem{hartmann16}
Lee {Hartmann}, Gregory {Herczeg}, and Nuria {Calvet}.
\newblock {Accretion onto Pre-Main-Sequence Stars}.
\newblock {\em \araa}, 54:135--180, Sep 2016.

\bibitem{lopezMartinez14}
Fatima {L{\'o}pez-Mart{\'\i}nez} and Ana~In{\'e}s {G{\'o}mez de Castro}.
\newblock {Constraints to the magnetospheric properties of T Tauri stars - I.
  The C II], Fe II] and Si II] ultraviolet features}.
\newblock {\em \mnras}, 442(4):2951--2962, Aug 2014.

\bibitem{najita00}
J.~R. {Najita}, S.~{Edwards}, G.~{Basri}, and J.~{Carr}.
\newblock {Spectroscopy of Inner Protoplanetary Disks and the Star-Disk
  Interface}.
\newblock In V.~{Mannings}, A.~P. {Boss}, and S.~S. {Russell}, editors, {\em
  Protostars and Planets IV}, page 457, May 2000.

\bibitem{gomezDeCastro16}
Ana~I. {G{\'o}mez de Castro}, Boris {Gaensicke}, Coralie {Neiner}, and
  Martin~A. {Barstow}.
\newblock {Reflections on the discovery space for a large ultraviolet-visible
  telescope: inputs from the European-led EUVO exercise}.
\newblock {\em Journal of Astronomical Telescopes, Instruments, and Systems},
  2:041215, Oct 2016.

\bibitem{kurosawa13}
Ryuichi {Kurosawa} and M.~M. {Romanova}.
\newblock {Spectral variability of classical T Tauri stars accreting in an
  unstable regime}.
\newblock {\em \mnras}, 431(3):2673--2689, May 2013.

\bibitem{alencar2018}
S.~H.~P. {Alencar}, J.~{Bouvier}, J.~F. {Donati}, E.~{Alecian}, C.~P. {Folsom},
  K.~{Grankin}, G.~A.~J. {Hussain}, C.~{Hill}, A.~M. {Cody}, A.~{Carmona},
  C.~{Dougados}, S.~G. {Gregory}, G.~{Herczeg}, F.~{M{\'e}nard}, C.~{Moutou},
  L.~{Malo}, M.~{Takami}, and {Matysse Collaboration}.
\newblock {Inner disk structure of the classical T Tauri star LkCa 15}.
\newblock {\em \aap}, 620:A195, Dec 2018.

\bibitem{gregory12}
S.~G. {Gregory}, J.~F. {Donati}, J.~{Morin}, G.~A.~J. {Hussain}, N.~J. {Mayne},
  L.~A. {Hillenbrand}, and M.~{Jardine}.
\newblock {Can We Predict the Global Magnetic Topology of a Pre-main-sequence
  Star from Its Position in the Hertzsprung-Russell Diagram?}
\newblock {\em \apj}, 755(2):97, Aug 2012.

\bibitem{hill19}
C.~A. {Hill}, C.~P. {Folsom}, J.~F. {Donati}, G.~J. {Herczeg}, G.~A.~J.
  {Hussain}, S.~H.~P. {Alencar}, S.~G. {Gregory}, and {Matysse Collaboration}.
\newblock {Magnetic topologies of young suns: the weak-line T Tauri stars TWA 6
  and TWA 8A}.
\newblock {\em \mnras}, 484(4):5810--5833, Apr 2019.

\bibitem{villebrun19}
F.~{Villebrun}, E.~{Alecian}, G.~{Hussain}, J.~{Bouvier}, C.~P. {Folsom},
  Y.~{Lebreton}, L.~{Amard}, C.~{Charbonnel}, F.~{Gallet}, L.~{Haemmerl{\'e}},
  T.~{B{\"o}hm}, C.~{Johns-Krull}, O.~{Kochukhov}, S.~C. {Marsden}, J.~{Morin},
  and P.~{Petit}.
\newblock {Magnetic fields of intermediate-mass T Tauri stars. I. Magnetic
  detections and fundamental stellar parameters}.
\newblock {\em \aap}, 622:A72, Feb 2019.

\bibitem{emeriauViard17}
Constance {Emeriau-Viard} and Allan~Sacha {Brun}.
\newblock {Origin and Evolution of Magnetic Field in PMS Stars: Influence of
  Rotation and Structural Changes}.
\newblock {\em \apj}, 846(1):8, Sep 2017.

\bibitem{zanni13}
C.~{Zanni} and J.~{Ferreira}.
\newblock {MHD simulations of accretion onto a dipolar magnetosphere. II.
  Magnetospheric ejections and stellar spin-down}.
\newblock {\em \aap}, 550:A99, Feb 2013.

\bibitem{bergin03}
Edwin {Bergin}, Nuria {Calvet}, Paola {D'Alessio}, and Gregory~J. {Herczeg}.
\newblock {The Effects of UV Continuum and Ly{\ensuremath{\alpha}} Radiation on
  the Chemical Equilibrium of T Tauri Disks}.
\newblock {\em \apjl}, 591(2):L159--L162, Jul 2003.

\bibitem{fraschetti18}
F.~{Fraschetti}, J.~J. {Drake}, O.~{Cohen}, and C.~{Garraffo}.
\newblock {Mottled Protoplanetary Disk Ionization by Magnetically Channeled T
  Tauri Star Energetic Particles}.
\newblock {\em \apj}, 853(2):112, Feb 2018.

\bibitem{alexander14}
R.~{Alexander}, I.~{Pascucci}, S.~{Andrews}, P.~{Armitage}, and L.~{Cieza}.
\newblock {The Dispersal of Protoplanetary Disks}.
\newblock In Henrik {Beuther}, Ralf~S. {Klessen}, Cornelis~P. {Dullemond}, and
  Thomas {Henning}, editors, {\em Protostars and Planets VI}, page 475, Jan
  2014.

\bibitem{salyk09}
C.~{Salyk}, G.~A. {Blake}, A.~C.~A. {Boogert}, and J.~M. {Brown}.
\newblock {High-resolution 5 {\ensuremath{\mu}}m Spectroscopy of Transitional
  Disks}.
\newblock {\em \apj}, 699(1):330--347, Jul 2009.

\bibitem{trilling02}
D.~E. {Trilling}, J.~I. {Lunine}, and W.~{Benz}.
\newblock {Orbital migration and the frequency of giant planet formation}.
\newblock {\em \aap}, 394:241--251, Oct 2002.

\bibitem{baruteau14}
C.~{Baruteau}, A.~{Crida}, S.~J. {Paardekooper}, F.~{Masset}, J.~{Guilet},
  B.~{Bitsch}, R.~{Nelson}, W.~{Kley}, and J.~{Papaloizou}.
\newblock {Planet-Disk Interactions and Early Evolution of Planetary Systems}.
\newblock In Henrik {Beuther}, Ralf~S. {Klessen}, Cornelis~P. {Dullemond}, and
  Thomas {Henning}, editors, {\em Protostars and Planets VI}, page 667, Jan
  2014.

\bibitem{dupree05}
A.~K. {Dupree}, A.~{Lobel}, P.~R. {Young}, T.~B. {Ake}, J.~L. {Linsky}, and
  S.~{Redfield}.
\newblock {A Far-Ultraviolet Spectroscopic Survey of Luminous Cool Stars}.
\newblock {\em \apj}, 622(1):629--652, Mar 2005.

\bibitem{perezMartinez11}
M.~Isabel {P{\'e}rez Mart{\'\i}nez}, K.~P. {Schr{\"o}der}, and M.~{Cuntz}.
\newblock {The basal chromospheric Mg II h+k flux of evolved stars: probing the
  energy dissipation of giant chromospheres}.
\newblock {\em \mnras}, 414(1):418--427, Jun 2011.

\bibitem{josselin07}
E.~{Josselin} and B.~{Plez}.
\newblock {Atmospheric dynamics and the mass loss process in red supergiant
  stars}.
\newblock {\em \aap}, 469(2):671--680, Jul 2007.

\bibitem{konstantinovaAntova14}
Renada {Konstantinova-Antova}, Michel {Auri{\`e}re}, Corinne {Charbonnel},
  Natalia {Drake}, Gregg {Wade}, Svetla {Tsvetkova}, Pascal {Petit},
  Klaus-Peter {Schr{\"o}der}, and Agnes {L{\`e}bre}.
\newblock {Magnetic fields in single late-type giants in the Solar vicinity:
  How common is magnetic activity on the giant branches?}
\newblock In Pascal {Petit}, Moira {Jardine}, and Hendrik~C. {Spruit}, editors,
  {\em Magnetic Fields throughout Stellar Evolution}, volume 302 of {\em IAU
  Symposium}, pages 373--376, Aug 2014.

\bibitem{auriere15}
M.~{Auri{\`e}re}, R.~{Konstantinova-Antova}, C.~{Charbonnel}, G.~A. {Wade},
  S.~{Tsvetkova}, P.~{Petit}, B.~{Dintrans}, N.~A. {Drake}, T.~{Decressin},
  N.~{Lagarde}, J.~F. {Donati}, T.~{Roudier}, F.~{Ligni{\`e}res}, K.~P.
  {Schr{\"o}der}, J.~D. {Landstreet}, A.~{L{\`e}bre}, W.~W. {Weiss}, and J.~P.
  {Zahn}.
\newblock {The magnetic fields at the surface of active single G-K giants}.
\newblock {\em \aap}, 574:A90, Feb 2015.

\bibitem{charbonnel17}
C.~{Charbonnel}, T.~{Decressin}, N.~{Lagarde}, F.~{Gallet}, A.~{Palacios},
  M.~{Auri{\`e}re}, R.~{Konstantinova-Antova}, S.~{Mathis}, R.~I. {Anderson},
  and B.~{Dintrans}.
\newblock {The magnetic strip(s) in the advanced phases of stellar evolution.
  Theoretical convective turnover timescale and Rossby number for low- and
  intermediate-mass stars up to the AGB at various metallicities}.
\newblock {\em \aap}, 605:A102, Sep 2017.

\bibitem{vlemmings17}
Wouter {Vlemmings}, Theo {Khouri}, Eamon {O'Gorman}, Elvire {De Beck},
  Elizabeth {Humphreys}, Boy {Lankhaar}, Matthias {Maercker}, Hans {Olofsson},
  Sofia {Ramstedt}, Daniel {Tafoya}, and Aki {Takigawa}.
\newblock {The shock-heated atmosphere of an asymptotic giant branch star
  resolved by ALMA}.
\newblock {\em Nature Astronomy}, 1:848--853, Oct 2017.

\bibitem{rau19}
Gioia {Rau}, Jr. {Montez}, Rodolfo, Kenneth {Carpenter}, Markus {Wittkowski},
  Sara {Bladh}, Margarita {Karovska}, Vladimir {Airapetian}, Tom {Ayres},
  Martha {Boyer}, Andrea {Chiavassa}, Geoffrey {Clayton}, William {Danchi},
  Orsola {De Marco}, Andrea~K. {Dupree}, Tomasz {Kaminski}, Joel~H. {Kastner},
  Franz {Kerschbaum}, Jeffrey {Linsky}, Bruno {Lopez}, John {Monnier}, Miguel
  {Montarg{\`e}s}, Krister {Nielsen}, Keiichi {Ohnaka}, Sofia {Ramstedt},
  Rachael {Roettenbacher}, Theo {ten Brummelaar}, Claudia {Paladini},
  Arkaprabha {Sarangi}, Gerard {van Belle}, and Paolo {Ventura}.
\newblock {Cool, evolved stars: results, challenges, and promises for the next
  decade}.
\newblock {\em \baas}, 51(3):241, May 2019.

\bibitem{lopezAriste18}
A.~{L{\'o}pez Ariste}, P.~{Mathias}, B.~{Tessore}, A.~{L{\`e}bre},
  M.~{Auri{\`e}re}, P.~{Petit}, N.~{Ikhenache}, E.~{Josselin}, J.~{Morin}, and
  M.~{Montarg{\`e}s}.
\newblock {Convective cells in Betelgeuse: imaging through spectropolarimetry}.
\newblock {\em \aap}, 620:A199, Dec 2018.

\bibitem{kervella15}
P.~{Kervella}, W.~{Homan}, A.~M.~S. {Richards}, L.~{Decin}, I.~{McDonald},
  M.~{Montarg{\`e}s}, and K.~{Ohnaka}.
\newblock {ALMA observations of the nearby AGB star L$_{2}$ Puppis. I. Mass of
  the central star and detection of a candidate planet}.
\newblock {\em \aap}, 596:A92, Dec 2016.

\bibitem{Ertley2018}
C.~{Ertley}, O.~{Siegmund}, J.~{Vallerga}, A.~{Tremsin}, N.~{Darling},
  J.~{Hull}, J.~{Tedesco}, T.~{Curtis}, and C.~{Paw U.}
\newblock {Microchannel plate detectors for future NASA UV observatories}.
\newblock In {\em \procspie}, volume 10699 of {\em Society of Photo-Optical
  Instrumentation Engineers (SPIE) Conference Series}, page 106993H, Jul 2018.

\bibitem{Fleming2017}
Brian {Fleming}, Manuel {Quijada}, John {Hennessy}, Arika {Egan}, Javier {Del
  Hoyo}, Brian~A. {Hicks}, James {Wiley}, Nicholas {Kruczek}, Nicholas
  {Erickson}, and Kevin {France}.
\newblock {Advanced environmentally resistant lithium fluoride mirror coatings
  for the next generation of broadband space observatories}.
\newblock {\em \ao}, 56(36):9941, Dec 2017.

\bibitem{nikzad2017}
Shouleh {Nikzad}, April~D. {Jewell}, Michael~E. {Hoenk}, Todd~J. {Jones}, John
  {Hennessy}, Tim {Goodsall}, Alexander~G. {Carver}, Charles {Shapiro},
  Samuel~R. {Cheng}, Erika~T. {Hamden}, Gillian {Kyne}, D.~Christopher
  {Martin}, David {Schiminovich}, Paul {Scowen}, Kevin {France}, Stephan
  {McCand liss}, and Roxana~E. {Lupu}.
\newblock {High-efficiency UV/optical/NIR detectors for large aperture
  telescopes and UV explorer missions: development of and field observations
  with delta-doped arrays}.
\newblock {\em Journal of Astronomical Telescopes, Instruments, and Systems},
  3:036002, Jul 2017.

\end{thebibliography}

\end{document}